\documentclass[a4paper,11pt]{article}

\usepackage{jheppub} 

\usepackage[T1]{fontenc} 
\usepackage{epstopdf}
\newcommand{\hs}{\hspace*{0.5cm}}

\newcommand{\be}{\begin{equation}}
\newcommand{\ee}{\end{equation}}
\newcommand{\bea}{\begin{eqnarray}}
\newcommand{\eea}{\end{eqnarray}}
\newcommand{\ben}{\begin{enumerate}}
\newcommand{\een}{\end{enumerate}}
\newcommand{\bde}{\begin{widetext}}
\newcommand{\ede}{\end{widetext}}
\newcommand{\nn}{\nonumber}
\newcommand{\crn}{\nonumber \\}

\newcommand{\al}{\alpha}
\newcommand{\la}{\lambda}

\newcommand{\ga}{\gamma}

\newcommand{\om}{\omega}

\newcommand{\fr}{\frac}
\newcommand{\bc}{\begin{center}}
\newcommand{\ec}{\end{center}}
\newcommand{\Ga}{\Gamma}
\newcommand{\de}{\delta}

\newcommand{\La}{\Lambda}
\newcommand{\si}{\sigma}

\newcommand{\bit}{\begin{itemize}}
\newcommand{\eit}{\end{itemize}}
\title{\boldmath The $T_7$ flavor symmetry in 3-3-1 model with neutral leptons}

\author[a]{V. V. Vien,}
\author[b]{H. N. Long}

\affiliation[a]{Department of Physics, Tay Nguyen University, 567 Le
Duan, Buon Ma Thuot, Vietnam}

\affiliation[b]{Institute of Physics,
VAST, 10 Dao Tan, Ba Dinh, Hanoi, Vietnam}
\emailAdd{wvienk16@gmail.com}
\emailAdd{hnlong@iop.vast.ac.vn}

\abstract{We construct a 3-3-1 model based on non-Abelian discrete
symmetry $T_7$ responsible for the fermion masses. Neutrinos get
masses from only anti-sextets which are in triplets
$\underline{3}$ and $\underline{3}^*$ under $T_7$. The flavor
mixing patterns and mass splitting are obtained without
perturbation. The tribimaximal form obtained with the breaking
$T_7 \rightarrow Z_3$ in charged lepton sector and both $T_7
\rightarrow Z_3$ and $Z_3 \rightarrow \{\mathrm{Identity}\}$ must
be taken place in neutrino sector but only apart in breakings $Z_3
\rightarrow \{\mathrm{Identity}\}$ (without contribution of
$\si'$), and the upper bound on neutrino mass $\sum_{i=1}^3m_i$ at
the level is presented. The Dirac CP violation phase $\delta$ is predicted to either
  $\frac{\pi}{2}$ or $\frac{3\pi}{2}$ which
 is maximal CP violation. From the Dirac CP violation phase we obtain the relation
  between Euler's angles which is consistent with the experimental in PDG
2012. On the other hand, the realistic lepton mixing can be
obtained if both the direction for breakings $T_7 \rightarrow Z_3$
and $Z_3 \rightarrow \{\mathrm{Identity}\}$ are taken place in
neutrino sectors. The CKM  matrix is the identity matrix at the
tree-level. }

\begin{document}
\maketitle
\flushbottom

\section{\label{intro}Introduction}
Despite the great success of the Standard Model (SM) of the elementary particle physics,
the origin of flavor structure, masses and mixings between generations of matter particles are unknown yet.
The neutrino mass and mixing is one of the most important evidence of beyond Standard Model physics. Many
experiments show that neutrinos have tiny masses and their mixing
is sill mysterious \cite{altar1, altar2}.

\textbf{The tri-bimaximal form for explaining the lepton mixing scheme was first} proposed by Harrison-Perkins-Scott (HPS), which
apart from the phase redefinitions, is given by \cite{hps1,hps2,hps3,hps4}
\begin{eqnarray}
U_{\mathrm{HPS}}=\left(
\begin{array}{ccc}
\frac{2}{\sqrt{6}}       &\frac{1}{\sqrt{3}}  &0\\
-\frac{1}{\sqrt{6}}      &\frac{1}{\sqrt{3}}  &\frac{1}{\sqrt{2}}\\
-\frac{1}{\sqrt{6}}      &\frac{1}{\sqrt{3}}  &-\frac{1}{\sqrt{2}}
\end{array}\right),\label{Uhps}
\end{eqnarray}
\textbf{can be considered  as a good approximation for the recent
neutrino experimental data.}

The most recent data are a clear sign of
rather large value $\theta_{13}$ \cite{smirnov} . The
data in PDG2012 \cite{PDG2012, PDG1, PDG2, PDG3, PDG4}  imply: \bea
&&\sin^2(2\theta_{12})=0.857\pm 0.024,\,\,\, \sin^2(2\theta_{13})=0.098\pm 0.013,\,\,\sin^2(2\theta_{23})> 0.95,\crn
&& \Delta m^2_{21}=(7.50\pm0.20)\times 10^{-5}
\mathrm{eV}^2,\,\, \Delta m^2_{32}=(2.32^{+0.12}_{-0.08})\times
10^{-3}\mathrm{eV}^2.\label{PDG2012}\eea
These large neutrino mixing angles are completely different from the quark
 mixing ones defined by the Cabibbo- Kobayashi-Maskawa (CKM) matrix \cite{CKM, CKM1}   .
 This has stimulated work on flavor symmetries and non-Abelian discrete
 symmetries are considered to be the most attractive candidate to
formulate dynamical principles that can lead to the flavor
mixing patterns for quarks and lepton. There are many recent models based on
the non-Abelian discrete symmetries, such as $A_4$ ~\cite{A41, A42, A43, A44, A45,A46, A47, A48, A49, A410, A411, A412, A413, A414, A415, A416, A417, A418, dlsh}  , $A_5$ \cite{A51, A52, A53, A54, A55, A56, A57, A58, A59, A510, A511, A512, A513} ,
$S_3$\cite{S31,S32,S33,S34,S35,S36,S37,S38,S39,S310,S311,S312,S313,S314,S315,S316,S317,S318,S319,S320,S321,S322,S323,S324,S325,S326,S327,S328,S329,S330,S331,S332,S333,S334,S335,S336,S337,S338,S339,S340,S341,S342} , $S_4$ \cite{S41,S42,S43,S44,S45,S46,S47,S48,S49,S410,S411,S412,S413,S414,S415,S416,S417,S418,S419,S420,S421,S422,S423,S424,S425,S426,S427,S428,S429} , $D_4$ \cite{D41,D42,D43,D44,D45,D46,D47,D48,D49,D410,D411,D412}, $D_5$ \cite{D51, D52}   ,
$T'$ \cite{Tp1,Tp2,Tp3,Tp4,Tp5,Tp6,Tp7,Tp8,Tp9,Tp10,Tp11,Tp12}, $T_7$ \cite{T71, T72, T73, T74, T75, T76, T77} and so forth.

Among the possible extensions of SM, a curious choice are the
3-3-1 models which encompass a class of models based on the gauge
group  $\mbox{SU}(3)_C\otimes \mbox{SU}(3)_L \otimes
\mbox{U}(1)_X$~ \cite{331m1,331m2,331m3,331m4,331m5, 331r1, 331r2,
331r3, 331r4,
 331r5, 331r6, e3311, e3312,e331v1, e331v2, e3313, e3314}, that is at first spontaneously broken to the SM group
$\mbox{SU}(3)_C\otimes \mbox{SU}(2)_L \otimes \mbox{U}(1)_Y$ and
then undergoes the spontaneously broken to $\mbox{SU}(3)_C \otimes
\mbox{U}(1)_Q$. The extension of the gauge group with respect to
SM leads to interesting consequences.
 The first one is that the requirement of
anomaly cancelation together with that of asymptotic freedom of
QCD implies that the number of generations must necessarily be
equal to the number of colors, hence giving an explanation for the
existence of three generations. Furthermore, quark generations
should transform differently under the action of $SU(3)_L$. In
particular, two quark generations  should transform as triplets,
one as an antitriplet.

A fundamental relation holds among some of the generators of the
group \textbf{\cite{T76, T77}}: \bea
Q=T_3+\beta T_8+X \label{Qoperator}\eea
where $Q$ indicates the electric
charge, $T_3$ and $T_8$ are two of the $SU(3)$ generators and $X$
is the generator of $U(1)_X$. $\beta$ is a key parameter that
defines a specific variant of the model.

The model thus provides a partial explanation of the family
number, as also required by flavor symmetries such as $T_7$ for
3-dimensional representations. In addition, due to the anomaly
cancelation one family of quarks has to transform under
$\mathrm{SU}(3)_L$ differently from the two others. $T_7$ can meet
this requirement with three inequivalent representations
 $\underline{1}, \underline{1}',\underline{1}''$.
Note that $T_7$ has not been considered before in the kind of the
3-3-1 model.

There are two typical variants of the 3-3-1 models as far as
lepton sectors are concerned. In the minimal version, three
$\mathrm{SU}(3)_L$ lepton triplets are $(\nu_L,l_L,l^c_R)$, where
$l_{R}$ are ordinary right-handed charged-leptons \cite{331m1,
331m2, 331m3, 331m4, 331m5}   . In the second version, the third
components of lepton triplets are the  right-handed neutrinos,
$(\nu_L,l_L,\nu^c_R)$ \cite{331r1,331r2,331r3,331r4,331r5,331r6}.
To have a model with the realistic neutrino mixing matrix, we
should consider another variant of the form $(\nu_L,l_L,N^c_R)$
where $N_R$ are three new fermion singlets under standard model
symmetry with vanishing lepton-numbers \cite{dlshA4, dlsvS4,
dlnvS3, vlD4}   .

In our previous works \cite{dlshA4, dlsvS4, dlnvS3, vlD4}   , the discrete symmetries
have been explored to the 3-3-1 models. In Ref. \cite{dlsvS4} we have studied the 3-3-1
model with neutral fermions based
    on $S_4$ group, in which most of the Higgs multiplets are in triplets under $S_4$
    except $\chi$ lying  in a singlet\footnote{\textbf{$\chi$ is the unique singlet under $S_4$ defined
    in expression (25) in Ref.\cite{dlsvS4}}}, and the exact tribimaximal form \cite{hps1,hps2,hps3,hps4} is obtained,
     where $\theta_{13}= 0$. As we know, the recent considerations have implied $\theta_{13}\neq 0$ \cite{A41,
      A42, A43, A44, A45,A46, A47, A48, A49, A410, A411, A412, A413, A414, A415, A416, A417, A418, S31,S32,S33,
      S34,S35,S36,S37,S38,S39,S310,S311,S312,S313,S314,S315,S316,S317,S318,S319,S320,S321,S322,S323,S324,S325,
      S326,S327,S328,S329,S330,S331,S332,S333,S334,S335,S336,S337,S338,S339,S340,S341,S342, S41,S42,S43,S44,S45,
      S46,S47,S48,S49,S410,S411,S412,S413,S414,S415,S416,S417,S418,S419,S420,S421,S422,S423,S424,S425,S426,S427,
      S428,S429}, but small as given in (\ref{PDG2012}).
     This problem has been improved in this type of the model in Ref. \cite{dlnvS3, vlD4} by adding new $SU(3)_L$
      multiplets and one of them is regarded as a small perturbation. The model therefore contains up to eight
      Higgs multiplets, and the scalar potential
of the model is quite complicated \cite{dlnvS3, vlD4}.

CP violation plays a crucial role in our understanding of the
observed baryon asymmetry of the Universe \cite{CPvio}. In the SM, CP symmetry is violated due
to a complex phase in the CKM matrix \cite{CKM, CKM1}. However,
since the extent
 of CP violation in the SM is not enough for achieving the observed BAU,  we need new source of CP violation for successful BAU.
On the other hand, CP violations in the lepton sector are
imperative if the BAU could be realized through leptogenesis. So,
any hint or observation of the leptonic CP violation  can
strengthen our belief in leptogenesis \cite{CPvio}.

The violation of the CP symmetry is a crucial ingredient of any
dynamical mechanism which intends to explain both low energy CP
violation and the baryon asymmetry. Renormalizable gauge theories
are based
 on the spontaneous symmetry breaking mechanism, and it is natural to have the spontaneous CP violation
  as an integral part of that mechanism. Determining all
possible sources of CP violation is a fundamental challenge for high energy physics. In theoretical
 and economical viewpoints, the spontaneous CP breaking necessary to generate the baryon asymmetry
  and leptonic CP violation at low energies brings us to a common source which comes from the
  phase of the scalar field responsible for the spontaneous CP breaking at a high energy scale \cite{CPvio}.

In this paper, we investigate another choice with $T_7$, the
smallest group with two non-equivalent 3-dimensional irreducible
representations, contains two triplet irreducible representations
and three singlets which play a crucial role in consistently
reproducing fermion masses and mixing. As we will see, $T_7$ model
has some new features since fewer Higgs multiplets are needed in
order to allow the fermions to gain masses and to break symmetries
and the physics we will see is different from the former. The CP
violation is the first time considered under $SU(3)_L\times
U(1)_X$ model based on $T_{7}$ flavor symmetry in which the
$T_{7}$ symmetry avoids the mass degeneracy of lepton masses. The
light neutrino masses can be generated at tree level, and the
vacuum alignment problem which arises in the presence of two
$T_{7}$-triplet scalar fields $\underline{3}, \underline{3}^*$ can
naturally explain the measured value of $\theta_{13}$ and thereby
the hierarchy of neutrino masses. The seesaw mechanism can explain
the smallness of the measured neutrino masses and the maximal
Dirac CP violation.

The rest of this work is organized
as follows. In Sec. \ref{fermion} and \ref{Chargedlep} we present
the necessary elements of the 3-3-1 model with the $T_7$ symmetry
as well as introducing necessary Higgs fields responsible for the
charged lepton masses. In Sec. \ref{quark}, we  discuss on quark
sector. Sec. \ref{neutrino} is devoted for the neutrino mass and
mixing. We summarize our results and make conclusions in the
section \ref{conclus}. Appendix \ref{apa} presents a brief of the
$T_7$ theory. Appendix \ref{apb} provides the lepton number ($L$)
and lepton parity ($P_l$) of  particles in the model.

\section{Fermion content\label{fermion}}

The gauge symmetry is based on $\mathrm{SU}(3)_C\otimes
\mathrm{SU}(3)_L \otimes \mathrm{U}(1)_X$, where the electroweak
factor $\mathrm{SU}(3)_L \otimes \mathrm{U}(1)_X$ is extended from
those of the Standard Model (SM), and the strong interaction sector is retained.
Each lepton family includes a new neutral fermion $(N_R)$ with vanishing
lepton number $L(N_R)=0$ arranged under the
$\mathrm{SU}(3)_L$ symmetry as a triplet $(\nu_L, l_L, N^c_R)$ and
a singlet $l_R$. The residual electric charge operator $Q$ is
therefore related to the generators of the gauge symmetry by \bea
Q=T_3-\fr{1}{\sqrt{3}}T_8+X, \label{Qnf}\eea
 where $T_a$ $(a=1,2,...,8)$ are
$\mathrm{SU}(3)_L$ charges with $\mathrm{Tr}T_aT_b=\fr 1 2
\de_{ab}$ and $X$ is the  $\mathrm{U}(1)_X$ charge. This means
that the model under consideration does not contain exotic
electric charges in the fundamental fermion, scalar and adjoint
gauge boson representations.

Since the particles in the lepton triplet have different lepton
number (1 and 0), so the lepton number in the model  does not
commute with the gauge symmetry unlike the SM. Therefore, it is
better to work with a new conserved charge $\mathcal{L}$ commuting
with the gauge symmetry and related to the ordinary lepton number
by diagonal matrices \cite{dlshA4, dlsvS4, dlnvS3, clong, vlD4}
 \bea L=\fr{2}{\sqrt{3}}T_8+\mathcal{L}.\label{Lnf}\eea

The lepton charge arranged in this way, i.e. $L(N_R)=0$, is in order to prevent unwanted interactions due to
$\mathrm{U}(1)_\mathcal{L}$ symmetry and breaking due to the
lepton parity to obtain the consistent lepton and
quark spectra. By this embedding, exotic quarks $U, D$ as well as
new non-Hermitian gauge bosons $X^0$, $Y^\pm$ possess lepton
charges as of the ordinary leptons:
$L(D)=-L(U)=L(X^0)=L(Y^{-})=1$.

Under the $[\mathrm{SU}(3)_L, \mathrm{U}(1)_X,
\mathrm{U}(1)_\mathcal{L},\underline{T}_7]$ symmetries as
proposed, the fermions of the model transform as follows \bea
\psi_{L} &\equiv& \psi_{1,2,3L}=\left( \nu_{L} \hs
    l_{L} \hs
    N^c_{R}\right)^T\sim [3,-1/3,2/3,\underline{3}],\crn
l_{1R}&\sim&[1,-1,1,\underline{1}],\hs
l_{2 R}\sim[1,-1,1,\underline{1}'], \hs
l_{3 R}\sim[1,-1,1,\underline{1}''],\crn
Q_{1 L}&\equiv &
 \left( d_{1 L} \hs
  -u_{1 L}  \hs
    D_{1 L}\right)^T\sim[3^*,0,1/3,\underline{1}'], \crn
Q_{2 L}&\equiv &
 \left( d_{2 L} \hs
  -u_{2 L}  \hs
    D_{2 L}\right)^T\sim[3^*,0,1/3,\underline{1}''],\label{Fermcont}\\
 Q_{3L}&=& \left(u_{3L} \hs
    d_{3L} \hs
    U_{L} \right)^T\sim[3,1/3,-1/3,\underline{1}],\crn
u_{R} &\sim &u_{1,2,3R}=[1,2/3,0,\underline{3}], \hs
d_{R}\sim[1,-1/3,0,\underline{3}^*],\crn
U_R&\sim&[1,2/3,-1,\underline{1}],\hs D_{1 R}
\sim[1,-1/3,1,\underline{1}''], \hs D_{2 R}
\sim[1,-1/3,1,\underline{1}'].\nn\eea where the subscript numbers
on field indicate to respective families which also  in order
define components of their $T_7$ multiplets. \textbf{$U$ and
$D_{1,2}$  are exotic quarks carrying lepton numbers $L(U)=-1$ and
$L(D_{1,2})=1$, known as leptoquarks. }In the following, we
consider possibilities of generating the masses for the fermions.
The scalar multiplets needed for the purpose are also introduced.

\section{\label{Chargedlep}Charged lepton masses}
    The charged lepton masses arise from the couplings of $\bar{\psi}_{L} l_{1R}, \bar{\psi}_{L} l_{2R}$
     and $\bar{\psi}_{L} l_{3R}$ to scalars, where $\bar{\psi}_{L} l_{iL}\, (i=1,2,3)$ transforms as $3^*$ under
$\mathrm{SU}(3)_L$ and $\underline{3}^*$  under
$T_7$. To generate masses for charged leptons, we need a $SU(3)_L$ Higgs triplets that lying
in $\underline{3}$ under $T_7$ and transforms as $3$ under $\mathrm{SU}(3)_L$,
\bea \phi_i = \left(%
\begin{array}{c}
  \phi^+_{i1} \\
  \phi^0_{i2} \\
  \phi^+_{i3} \\
\end{array}%
\right)\sim [3,2/3,-1/3, \underline{3}] \hs (i=1,2,3) \label{phi}.\eea
Following the potential minimization
conditions, we have the followings alignments:
\begin{itemize}
\item[(1)] The first alignment: $\langle \phi_1\rangle= \langle \phi_2\rangle=\langle \phi_3\rangle$
then $T_7$ is broken into $Z_3$ consisting of the elements \{$e, b, b^2$\}.

\item[(2)] The second alignment: $\langle \phi_1\rangle\neq \langle \phi_2\rangle\neq\langle \phi_3\rangle$ or $\langle \phi_1\rangle\neq \langle \phi_2\rangle=\langle \phi_3\rangle$ or $\langle \phi_2\rangle\neq \langle \phi_1\rangle\neq\langle \phi_3\rangle$ or $\langle \phi_3\rangle\neq \langle \phi_1\rangle\neq\langle \phi_2\rangle$ then $T_7$ is broken into $\{\mathrm{Identity}\}$.

\item[(3)] The third alignment: $0=\langle \phi_1\rangle\neq\langle \phi_2\rangle=\langle \phi_3\rangle \neq 0$ or $0=\langle \phi_2\rangle\neq\langle \phi_3\rangle=\langle \phi_1\rangle\neq 0$ or $0=\langle \phi_3\rangle\neq\langle \phi_1\rangle=\langle \phi_2\rangle\neq 0$ then $T_7$ is broken into $\{\mathrm{Identity}\}$.

\item[(4)] The fourth alignment:  $0=\langle \phi_1\rangle\neq\langle \phi_2\rangle\neq\langle \phi_3\rangle\neq0$ or $0=\langle \phi_2\rangle\neq\langle \phi_1\rangle\neq\langle \phi_3\rangle\neq 0$ or $0=\langle \phi_3\rangle\neq\langle \phi_2\rangle\neq\langle \phi_1\rangle\neq0$ then $T_7$ is broken into $\{\mathrm{Identity}\}$.

\item[(5)] The fifth alignment: $0=\langle \phi_1\rangle=\langle \phi_2\rangle\neq\langle \phi_3\rangle\neq0$ or $0=\langle \phi_1\rangle=\langle \phi_3\rangle\neq\langle \phi_2\rangle\neq0$ or $0=\langle \phi_2\rangle=\langle \phi_3\rangle\neq\langle \phi_1\rangle\neq0$ then $T_7$ is broken into $\{\mathrm{Identity}\}$.
\end{itemize}
In this work, we
argue that only the first  alignment of VEV in charged - lepton sector is taken place, i.e, $T_7\rightarrow Z_3$, and this can be achieved by the Higgs triplet $\phi$ with the VEV alignment $\langle \phi\rangle=(\langle \phi_1\rangle, \langle \phi_1\rangle, \langle \phi_1\rangle)$ under $T_7$, where
\be \langle \phi_1 \rangle = \left(0 \hs v \hs 0 \right)^T.\label{vevphi} \ee
The Yukawa interactions are
 \bea -\mathcal{L}_{l}&=&h_1 (\bar{\psi}_{L}\phi)_{\underline{1}} l_{1R}+h_2 (\bar{\psi}_{L}\phi)_{\underline{1}''}l_{2R}
+h_3 (\bar{\psi}_{i L}\phi)_{\underline{1}'} l_{3R}+H.c\crn
&=&h_1
(\bar{\psi}_{1L}\phi_1+\bar{\psi}_{2L}\phi_2+\bar{\psi}_{3L}\phi_3)l_{1R}\crn
&+&h_2 (\bar{\psi}_{1L}\phi_1+\om^2\bar{\psi}_{2L}\phi_2+\om
\bar{\psi}_{3L}\phi_3)l_{2R}\crn &+&h_3
(\bar{\psi}_{1L}\phi_1+\om\bar{\psi}_{2L}\phi_2+\om^2
\bar{\psi}_{3L}\phi_3) l_{3R}+H.c.\eea
The mass Lagrangian for the charged leptons is then given by
\bea
-\mathcal{L}^{\mathrm{mass}}_l&=&h_1 v\bar{l}_{1L} l_{1R}+h_2 v\bar{l}_{1L} l_{2R}+h_3 v\bar{l}_{1L} l_{3R}\crn
&+&h_1v\bar{l}_{2L} l_{1R}+h_2v \om^2\bar{l}_{2L} l_{2R}+h_3v \om\bar{l}_{2L} l_{3R}\crn
&+&h_1v\bar{l}_{3L} l_{1R}+h_2v\om \bar{l}_{3L} l_{2R}+h_3v\om^2 \bar{l}_{3L} l_{3R}+H.c. \eea
The mass
Lagrangian for the charged leptons reads
\bea
-\mathcal{L}^{\mathrm{mass}}_l=(\bar{l}_{1L},\bar{l}_{2L},\bar{l}_{3L})
M_l (l_{1R},l_{2R},l_{3R})^T+H.c, \eea
where \be M_l=
\left(%
\begin{array}{ccc}
  h_1v&\,\,\, h_2v &\,\,\, h_3 v \\
   h_1v & \,\,\,\,\,\,\,h_2 v\om^2 &\,\,\,\, h_3 v \om \\
  h_1v & \,\,\,\,\, h_2 v \om &\,\,\,\,\,\,h_3 v \om^2\\
\end{array}%
\right).\label{Mltq}\ee
This matrix can be diagonalized as,
\bea U^{\dagger}_L M_lU_R=\left(%
\begin{array}{ccc}
  \sqrt{3}h_1 v & 0 & 0 \\
  0 & \sqrt{3}h_2 v & 0 \\
  0 & 0 & \sqrt{3}h_3 v \\
\end{array}%
\right)\equiv \left(%
\begin{array}{ccc}
  m_e & 0 & 0 \\
  0 & m_\mu & 0 \\
  0 & 0 & m_\tau \\
\end{array}%
\right),\label{Mld}\eea where \bea U_L=\fr{1}{\sqrt{3}}\left(%
\begin{array}{ccc}
  1 &\,\,\, 1 &\,\,\, 1 \\
  1 &\,\,\, \om^2 &\,\,\, \om \\
  1 &\,\,\, \om &\,\,\, \om^2 \\
\end{array}%
\right),\hs U_R=1.\label{Uclep}\eea
As will see in section \ref{neutrino}, in this case, the exact tribimaximal
mixing form is obtained, by choosing the right vev's in the neutrino sector.

The experimental  values for masses of  the charged leptons at the weak
scale are given as \cite{PDG2012}  : \bea m_e=0.511\, \textrm{MeV},\hs \
 m_{\mu}=105.658 \ \textrm{MeV},\hs m_{\tau}=1776.82\,
\textrm{MeV} \eea from which it follows that $h_1\ll h_2\ll h_3$.
On the other hand, if we choose the VEV $v=100 GeV$ then $h_1
\sim 10^{-6},\, h_2\sim 10^{-4},\, h_3\sim 10^{-2}$.

\section{\label{quark}Quark masses}
\begin{table}
\vspace*{-0.5cm}
\bc
\caption{List of couplings which form a singlet from the invariance under the $T_7$ \label{quarkcoup}}
\begin{tabular}{|c|c|c|c|c|}
\hline Couplings& Higgs multiplets\\ \hline\hline
$\bar{Q}_{3L}U_{R}\sim \left(3^*, \frac{1}{3}, -\frac{2}{3}, \underline{1}\right)$&$\chi\sim \left(3, -\frac{1}{3}, \frac{2}{3}, \underline{1}\right)$\\\hline
$\bar{Q}_{2 L}D_{2 R}\sim\left(3, -\frac{1}{3},\frac{2}{3}, \underline{1}\right)$ &$\chi^*\sim \left(3^*, \frac{1}{3}, -\frac{2}{3}, \underline{1}\right)$
\\\hline
$\bar{Q}_{1 L}D_{1R}\sim\left(3, -\frac{1}{3},\frac{2}{3}, \underline{1}\right)$ &$\chi^*\sim \left(3^*, \frac{1}{3}, -\frac{2}{3}, \underline{1}\right)$
\\\hline\hline
 $\bar{Q}_{3L}d_{R}\sim \left(3^*, -\frac{2}{3}, \frac{1}{3}, \underline{3}^*\right)$&  $\phi\sim \left(3, \frac{2}{3}, -\frac{1}{3},\underline{3}\right)$    \\\hline
 $\bar{Q}_{1 L}u_{R}\sim \left(3, \frac{2}{3}, -\frac{1}{3}, \underline{3}\right) $ & $\phi^*\sim \left(3^*, -\frac{2}{3}, \frac{1}{3}, \underline{3}^*\right)$
 \\\hline
 $\bar{Q}_{2 L}u_{R}\sim \left(3, \frac{2}{3}, -\frac{1}{3}, \underline{3}\right) $ & $\phi^*\sim \left(3^*, -\frac{2}{3}, \frac{1}{3}, \underline{3}^*\right)$
 \\\hline\hline
 $\bar{Q}_{3L}u_{R}\sim  \left(3^*, \frac{1}{3}, \frac{1}{3},\underline{3}\right)$&$\eta\sim \left(3, -\frac{1}{3}, -\frac{1}{3}, \underline{3}^*\right)$
  \\\hline
 $\bar{Q}_{1L}d_{R}\sim  \left(3, -\frac{1}{3}, -\frac{1}{3}, \underline{3}^*\right)$&$\eta^*\sim \left(3^*, \frac{1}{3}, \frac{1}{3}, \underline{3}\right)$
  \\\hline
 $\bar{Q}_{2L}d_{R}\sim  \left(3, -\frac{1}{3}, -\frac{1}{3}, \underline{3}^*\right)$&$\eta^*\sim \left(3^*, \frac{1}{3}, \frac{1}{3}, \underline{3}\right)$ \\\hline
 \end{tabular}
\ec
\vspace*{-0.5cm}
\end{table}

To generate masses for quarks with a minimal Higgs content, we
additionally introduce the following Higgs triplets:
\bea \eta_i&=&
\left(%
\begin{array}{c}
  \eta^0_{i1} \\
  \eta^-_{i2} \\
  \eta^0_{i3} \\
\end{array}%
\right)\sim \left[3,-1/3,-1/3, \underline{3}\right]\hs  (i=1,2,3),\\
\chi&=&\left(%
\begin{array}{c}
  \chi^0_1 \\
  \chi^-_2 \\
  \chi^0_3 \\
\end{array}%
\right)\sim \left[3,-1/3,2/3,\underline{1}\right].\eea
The
Higgs content and Yukawa couplings in the quark sector are
summarized in Table \ref{quarkcoup}.

The Yukawa
interactions are \bea -\mathcal{L}_q &=& h^d_3 \bar{Q}_{3L}(\phi
d_R)_1 + h^u_1 \bar{Q}_{1L}(\phi^*u_R)_{1''}+ h^u_2
\bar{Q}_{2L}(\phi^*u_R)_{1'}\crn
&+& h^u_3 \bar{Q}_{3L}(\eta
u_R)_1+h^d_1 \bar{Q}_{1L}(\eta^* d_R)_{1''}+h^d_2
\bar{Q}_{2L}(\eta^* d_R)_{1'}\crn
&+&  f_3 \bar{Q}_{3L}\chi U_R +
f_1 \bar{Q}_{1L}\chi^* D_{1R}+f_2 \bar{Q}_{2L}\chi^* D_{2R}+H.c
\crn &=& h^d_3 \bar{Q}_{3L}(\phi_1 d_{1R}+\phi_2
d_{2R}+\phi_3 d_{3R}) \crn &+& h^u_1
\bar{Q}_{1L}(\phi^*_1u_{R}+\om^2\phi^*_2u_{2R}+\om\phi^*_3u_{3R})\crn
&+&h^u_2
\bar{Q}_{2L}(\phi^*_1u_{R}+\om\phi^*_2u_{2R}+\om^2\phi^*_3u_{3R})\crn
&+& h^u_3 \bar{Q}_{3L}(\eta_1 u_{1R}+\eta_2 u_{2R}+\eta_3
u_{3R})\crn &+&h^d_1 \bar{Q}_{1L}(\eta^*_1 d_{1R}+\om^2\eta^*_2
d_{2R}+\om\eta^*_3 d_{3R})\crn &+&h^d_2 \bar{Q}_{2L}(\eta^*_1
d_{1R}+\om\eta^*_2 d_{2R}+\om^2\eta^*_3 d_{3R})\crn &+&  f_3
\bar{Q}_{3L}\chi U_R + f_1 \bar{Q}_{1L}\chi^* D_{1R}+f_2
\bar{Q}_{2L}\chi^* D_{2R}+H.c.\eea
We suppose that $T_7$ is broken into $Z_3$ like the case of the charged lepton
sector, i,e, the VEVs of $\eta$ and $\chi$ \textbf{are given as} $\langle\eta\rangle=
(\langle\eta_1\rangle, \langle\eta_1\rangle,\langle\eta_1\rangle)$ with
\bea
 \langle\eta_1\rangle&=&
\left( u \hs   0 \hs   0\right)^T, \label{veveta}
\eea
\textbf{and}
\bea
 \langle\chi\rangle&=&
\left( 0 \hs   0 \hs   v_\chi\right)^T. \label{vevchi}
\eea
The mass Lagrangian for quarks is given by
\bea -\mathcal{L}^{mass}_q &=&- h^u_1v_1\bar{u}_{1L}u_{1R}-h^u_1v_2
\om^2\bar{u}_{1L}u_{2R}- h^u_1 v_3\om\bar{u}_{1L}u_{3R}\crn
&-& h^u_2v_1\bar{u}_{2L}u_{1R}-h^u_2v_2\om\bar{u}_{2L}u_{2R}- h^u_2 v_3\om^2\bar{u}_{2L}u_{3R}\crn
&+& h^u_3u_1\bar{u}_{3L}u_{1R}+h^u_3u_2\bar{u}_{3L}u_{2R}+h^u_3u_3\bar{u}_{3L}u_{3R}\crn
&+& h^d_1u_1\bar{d}_{1L}d_{1R}+\om^2h^d_1u_2\bar{d}_{1L}d_{2R}+\om h^d_1u_3\bar{d}_{1L}d_{3R}\crn
&+& h^d_2u_1\bar{d}_{2L}d_{1R}+\om h^d_2u_2\bar{d}_{2L}d_{2R}+\om^2 h^d_2u_3\bar{d}_{2L}d_{3R}\crn
&+&h^d_3v_1 \bar{d}_{3L}
d_{1R}+h^d_3 v_2\bar{d}_{3L}d_{2R}+h^d_3 v_3\bar{d}_{3L}d_{3R}\crn
&+&  f_3 v_\chi\bar{U}_{L} U_R +
f_1 v_\chi\bar{D}_{1L}D_{1R}+f_2 v_\chi \bar{D}_{2L}D_{2R}+H.c.\eea
The exotic quarks get masses
\bea
m_U=f_3 v_\chi ,\hs m_{D_{1,2}}=f_{1,2}
v_\chi. \eea
The mass matrices for ordinary up-quarks and
down-quarks are, respectively, obtained as follows: \bea M_u =
\left(%
\begin{array}{ccc}
  -h^u_1 v &\,\,\, -h^u_1 v \om^2  & -h^u_1 v \om  \\
   -h^u_2 v &\,\, -h^u_2v \om   &\, -h^u_2  v \om^2   \\
  \,\,\,\, h^u_3 u &\,\,\, h^u_3 u & h^u_3 u \\
\end{array}%
\right),\,\,\, M_d=
\left(%
\begin{array}{ccc}
  h^d_1 u &\,\,\,\,\, h^d_1  u \om^2 &\,\,\, h^d_1  u \om  \\
   h^d_2 u &\,\,\,\, h^d_2 u \om  &\,\,\,\,\, h^d_2  u  \om^2 \\
  h^d_3 v & h^d_3 v & h^d_3 v \\
\end{array}%
\right).\label{MuMd} \eea \textbf{The structure of the up- and
down-quark mass matrices in (\ref{MuMd}) is similar to those in
\cite{A419}, i.e, in the model under consideration there is no CP
violation in the quark mixing matrix.}
 The mass matrices $M_u, M_d$ in (\ref{MuMd}) are diagonalized as follows
 \bea U^{u+}_LM_u U^{u}_R&=&
\left(%
\begin{array}{ccc}
  -\sqrt{3}h^u_1 v & 0 & 0 \\
  0 & -\sqrt{3}h^u_2 v & 0 \\
  0 & 0 & \sqrt{3}h^u_3 u \\
\end{array}%
\right)=\left(%
\begin{array}{ccc}
  m_u & 0 & 0 \\
  0 & m_c & 0 \\
  0 & 0 & m_t \\
\end{array}%
\right), \crn U^{d+}_LM_d U^{d}_R &=&
\left(%
\begin{array}{ccc}
  \sqrt{3}h^d_1 u& 0 & 0 \\
  0 & \sqrt{3}h^d_2 u & 0 \\
  0 & 0 & \sqrt{3}h^d_3 v \\
\end{array}%
\right)
=\left(%
\begin{array}{ccc}
  m_d & 0 & 0 \\
  0 & m_s & 0 \\
  0 & 0 & m_b \\
\end{array}%
\right).\label{quarkmasses}\eea
where \bea U^{u}_R=U^{d}_R=
\fr{1}{\sqrt{3}}\left(%
\begin{array}{ccc}
  1 & 1 & 1 \\
  \om & \om^2 & 1 \\
  \om^2 & \om & 1 \\
\end{array}%
\right),\hs U^u_L=U^d_L=1\label{UudLR}\eea
The unitary matrices, which couple the left-handed
up- and down-quarks to those in the mass bases are unit matrices. Therefore we get the
Cabibbo-Kobayashi-Maskawa (CKM) matrix \bea
U_\mathrm{CKM}=U^{d\dagger}_L U^u_L=1.\label{a41}\eea Note that
the property in (\ref{a41}) is common for some models based on the
discrete symmetry groups \cite{dlsvS4, dlshA4}.

The up and down quark masses are \bea m_u&=& -\sqrt{3}h^u_1 v,\hs
m_c= -\sqrt{3}h^u_2 v,\hs m_t =\sqrt{3}h^u_3 u,\crn
m_d&=&\sqrt{3}h^d_1 u,\hs\,\,\,\,\, m_s=\sqrt{3}h^d_2
u,\hs\,\,\,\,\, m_d =\sqrt{3}h^d_3 v. \eea
The current mass
values for the quarks are given by \cite{PDG2012}: \bea
m_u&=&2.3^{+0.7}_{-0.5}\ \textrm{MeV},\,\, m_c=1.275\pm 0.025\
\textrm{GeV},\,\, m_t=173.5\pm0.6\pm0.8,\ \textrm{GeV}\crn
m_d&=&4.8^{+0.7}_{-0.3}\ \textrm{MeV}, \,\, m_s=95\pm5\
\textrm{MeV},\hs\hs\,\,\, m_b=4.18\pm 0.03\
\textrm{GeV}.\label{vien3}\eea

It is obvious that if $|u| \sim|v|$, the
Yukawa coupling hierarchies are $|h^u_1| \ll h^u_2\ll h^u_3$, $h^d_1 \ll h^d_2\ll h^d_3$
and the couplings between up-quarks
$(h^u_2, h^u_3)$ and Higgs scalar multiplets are slightly
heavier than those of down-quarks $(h^d_2\ll h^d_3)$,
respectively.

 \section{\label{neutrino}Neutrino mass and mixing}

The neutrino masses arise from the couplings of $\bar{\psi}^c_{L} \psi_{L}$ to scalars,
 where $\bar{\psi}^c_{L} \psi_{L}$ transforms as $3^*\oplus 6$ under
$\mathrm{SU}(3)_L$ and $\underline{3}\oplus \underline{3}^*\oplus
\underline{3}^*$  under $T_7$. It is worth mentioning that, with
the $T_7$ group, $\underline{3} \times \underline{3} \times
\underline{3}$ has two invariants and $\underline{3} \times
\underline{3} \times \underline{3}^*$ has one invariant. For the
known scalar triplets $(\phi,\chi,\eta)$, there is no available
interaction because of the $\mathcal{L}$-symmetry. We will
therefore propose new SU(3)$_L$ anti-sextets instead coupling to
$\bar{\psi}^c_{ L}\psi_{ L}$ responsible for the neutrino masses.
To obtain a realistic neutrino spectrum, the antisextets transform
as follows\footnote{\textbf{ Note that in the model under
consideration, if the choice is a $SU(3)_L$ triplet [for example
$\rho\sim (3, 2/3, -4/3, \underline{3}^*)$] instead of $SU(3)_L$
anti-sextet $\si$ as in (\ref{sigma}), there will be a
contribution from term of $(\bar{\psi}^c_{ L}\psi_{
L})_{\underline{3}^*}\rho$ added to the elements $(M_D)_{11},
(M_D)_{22}, (M_D)_{33}$ of the matrix $M_D$ which is the same
order. The lepton mixing matrix therefore can only reach to
$U_{HPS}$} but not $\theta_{13}\neq0$}
 \be \sigma_i=
\left(%
\begin{array}{ccc}
  \sigma^0_{11} & \sigma^+_{12} & \sigma^0_{13} \\
  \sigma^+_{12} & \sigma^{++}_{22} & \sigma^+_{23} \\
  \sigma^0_{13} & \sigma^+_{23} & \sigma^0_{33} \\
\end{array}%
\right)_i\sim [6^*,2/3,-4/3,\underline{3}^*] \hs (i=1,2,3),\label{sigma} \ee

Following the potential minimization
conditions, we have the followings alignments:
\begin{itemize}
\item[(1)] The first alignment: $\langle \si_1\rangle= \langle \si_2\rangle=\langle \si_3\rangle$ then $T_7$
 is broken into $Z_3$   consisting of the elements \{$e, b, b^2$\}.

\item[(2)] The second alignment: $\langle \si_1\rangle\neq \langle \si_2\rangle\neq\langle \si_3\rangle$
or $\langle \si_1\rangle\neq \langle \si_2\rangle=\langle \si_3\rangle$ or $\langle \si_2\rangle\neq
\langle \si_1\rangle\neq\langle \si_3\rangle$ or $\langle \si_3\rangle\neq \langle \si_1
\rangle\neq\langle \si_2\rangle$ then $T_7$ is broken into $\{\mathrm{Identity}\}$.

\item[(3)] The third alignment: $0=\langle \si_1\rangle\neq\langle \si_2\rangle=\langle \si_3\rangle
\neq 0$ or $0=\langle \si_2\rangle\neq\langle \si_3\rangle=\langle \si_1\rangle \neq 0$
or $0=\langle \si_3\rangle\neq\langle \si_1\rangle=\langle \si_2\rangle\neq 0$  then $T_7$ is broken into $\{\mathrm{Identity}\}$.

\item[(4)] The fourth alignment: $0=\langle \si_1\rangle\neq\langle \si_2\rangle\neq\langle
\si_3\rangle\neq 0$ or $0= \langle \si_2\rangle\neq\langle \si_1\rangle\neq\langle \si_3
\rangle\neq 0$ or $0=\langle \si_3\rangle\neq\langle \si_1\rangle\neq\langle \si_2\rangle
\neq0$  then $T'$ is broken into $\{\mathrm{Identity}\}$.

\item[(5)] The fifth alignment: $0=\langle \si_1\rangle=\langle \si_2\rangle\neq\langle \si_3
\rangle\neq0$ or $0=\langle \si_1\rangle=\langle \si_3\rangle\neq\langle \si_2\rangle\neq0$
or $0=\langle \si_2\rangle=\langle \si_3\rangle\neq\langle \si_1\rangle\neq0$ then $T_7$ is broken into $\{\mathrm{Identity}\}$.
\end{itemize}
 To obtain a realistic neutrino spectrum, in this work we
argue that both the breakings $T_7\rightarrow Z_3$ and
$T_7\rightarrow \{\mathrm{identity}\}$ (Instead of $Z_3\rightarrow \{\mathrm{identity}\}$)
 must be taken place in neutrino sector. However, the VEVs of $\si$
does only one of these tasks. The $T_7\rightarrow Z_3$ can be achieved by a $SU(3)_L$
 anti-sextet $\si$ given in (\ref{sigma}) with the VEVs is set as $\langle \si\rangle=(\langle
 \si_1\rangle, \langle \si_1\rangle, \langle \si_1\rangle)$ under $T_7$, where
\bea \langle \si_1\rangle=\left(%
\begin{array}{ccc}
  \la_{\si} & 0 & v_{\si} \\
  0 & 0 & 0 \\
  v_{\si} & 0 & \La_{\si} \\
\end{array}%
\right). \label{sivev} \eea To achieve the second direction of the
breakings $T_7\rightarrow \{\mathrm{Identity}\}$
\textbf{(equivalently to $Z_3\rightarrow \{\mathrm{Identity}\}$)},
we additionally introduce another $SU(3)_L$ anti-sextet Higgs
scalar which is either put in  $\underline{3}$ or
$\underline{3}^*$ under $T_7$. This is equivalent to breaking the
subgroup $Z_3$ of the first direction into
$\{\mathrm{Identity}\}$, and it can be achieved within each case
below. \ben
\item A new $SU(3)_L$ anti-sextet $s$ which is put in the $\underline{3}$ under $T_7$,
 \bea s_i =
\left(%
\begin{array}{ccc}
  s^0_{11} & s^+_{12} & s^0_{13} \\
  s^+_{12} & s^{++}_{22} & s^+_{23} \\
  s^0_{13} & s^+_{23} & s^0_{33} \\
\end{array}%
\right)_i \sim [6^*,2/3,-4/3,\underline{3}],\label{s}\eea
with the VEVs given
by $\langle s \rangle = (\langle s_1 \rangle, 0,0)^T$, where
\bea \langle s_1\rangle=\left(%
\begin{array}{ccc}
  \la_{s} & 0 & v_{s} \\
  0 & 0 & 0 \\
  v_{s} & 0 & \La_{s} \\
\end{array}%
\right). \label{svev} \eea
\item Another $\mathrm{SU}(3)_L$ anti-sextet $\si'$ is put in the $\underline{3}^*$ under
$T_7$, with the VEVs chosen by
\bea  \sigma'&=&
\left(%
\begin{array}{ccc}
  \sigma'^0_{11} & \sigma'^+_{12} & \sigma'^0_{13} \\
  \sigma'^+_{12} & \sigma'^{++}_{22} & \sigma'^+_{23} \\
  \sigma'^0_{13} & \sigma'^+_{23} & \sigma'^0_{33} \\
\end{array}%
\right)\sim [6^*,2/3,-4/3,\underline{3}^*], \crn
\langle \si'_1\rangle&=&\left(%
\begin{array}{ccc}
  \la'_{\si} & 0 & v'_{\si} \\
  0 & 0 & 0 \\
  v'_{\si} & 0 & \La'_{\si} \\
\end{array}%
\right), \, \langle \si'_2\rangle =\langle \si'_3\rangle=0.
\label{sipvev} \eea \een Note that $\si'$ differs from $\si$ only
in the VEVs alignment. Combining both cases, after calculation, we
obtain the Yukawa interactions: \bea -\mathcal{L}_\nu&=&\fr 1 2 x
(\bar{\psi}^c_L \si)_{{\underline{3}}^*}\psi_L+y (\bar{\psi}^c_L
s)_{{\underline{3}}^*}\psi_L +\fr z 2(\bar{\psi}^c_L
\si')_{{\underline{3}}^*}\psi_L+H.c.\crn &=& \fr 1 2 x
(\bar{\psi}^c_{1L}\si_2\psi_{1L}+\bar{\psi}^c_{2L}\si_3
\psi_{2L}+\bar{\psi}^c_{3L}\si_1\psi_{3L})\crn &+&
y(\bar{\psi}^c_{2L}s_3\psi_{1L}+\bar{\psi}^c_{3L}s_1
\psi_{2L}+\bar{\psi}^c_{1L}s_2\psi_{3L})\crn &+&\fr z 2
(\bar{\psi}^c_{1L}\si'_2\psi_{1L}+\bar{\psi}^c_{2L}\si'_3
\psi_{2L}+\bar{\psi}^c_{3L}\si'_1\psi_{3L})
+H.c.\label{yn}\eea

The mass Lagrangian for the neutrinos is given by
\bea -\mathcal{L}^{\mathrm{mass}}_\nu &=&\fr 1 2 x(\la_{\si}\bar{\nu}^c_{1 L}\nu_{1L}+ v_{\si}\bar{\nu}^c_{1
L}N^c_{1R}+v_{\si}\bar{N}_{1R}\nu_{1L}+\La_{\si}\bar{N}_{1R}N^c_{1R})\crn
&+&\fr 1 2 x(\la_{\si}\bar{\nu}^c_{2 L}\nu_{2L}+v_{\si}\bar{\nu}^c_{2L}N^c_{2R}+
v_{\si}\bar{N}_{2R}\nu_{2L}+\La_{\si}\bar{N}_{2R}N^c_{2R})\crn
&+& \fr 1 2 x (\la_\si\bar{\nu}^c_{3 L}\nu_{3L}+v_\si\bar{\nu}^c_{3
L}N^c_{3R}+
v_\si\bar{N}_{3R}\nu_{3L}+\La_\si\bar{N}_{3R}N^c_{3R})\crn
&+&y(\la_{s}\bar{\nu}^c_{3 L}\nu_{2L}+v_{s}\bar{\nu}^c_{3 L}N^c_{2R}+
v_{s}\bar{N}_{3R}\nu_{2L}+\La_{s}\bar{N}_{3R}N^c_{2R})\crn
&+&\fr
1 2 z(\la'_\si\bar{\nu}^c_{3 L}\nu_{3L}+v'_\si\bar{\nu}^c_{3
L}N^c_{3R}+
v'_\si\bar{N}_{3R}\nu_{3L}+\La'_\si\bar{N}_{3R}N^c_{3R})
+H.c. \label{T7nm}\eea

The neutrino mass Lagrangian  can be written in matrix form as follows
 \be -\mathcal{L}^{\mathrm{mass}}_\nu=\fr 1 2
\bar{\chi}^c_L M_\nu \chi_L+ h.c.,\label{nm}\ee where
\bea \chi_L&\equiv&
\left(\nu_L \hs
  N^c_R \right)^T,\hs\,\,\, M_\nu\equiv\left(%
\begin{array}{cc}
  M_L & M^T_D \\
  M_D & M_R \\
\end{array}%
\right), \crn
\nu_L&=&(\nu_{1L},\nu_{2L},\nu_{3L})^T,\,\,
N_R=(N_{1R},N_{2R},N_{3R})^T, \eea
and the mass matrices are then obtained by
\be
M_{L,R,D}=\left(%
\begin{array}{ccc}
  a_{L,R,D} &\hs 0 &\hs 0 \\
  0 &\hs a_{L,R,D} &\hs b_{L,R,D} \\
 0 &\hs b_{L,R,D} &\hs a_{L,R,D}+c_{L,R,D} \\
\end{array}%
\right),\label{abcd}\ee
with
\bea
  a_{L} & =&\la_\si x, \hs  a_{D} =v_\si x, \hs  a_{R} =\La_\si x, \crn
  b_{L} & =&\la_sy ,\hs  b_{D}=v_sy,\hs  b_{R} =\La_sy,\crn
  c_{L} & =&\la'_\si z, \hs  c_{D} =v'_\si z, \hs  c_{R} =\La'_\si z. \label{bL,cL}\eea
Three observed neutrinos gain masses via a combination of type I
and type II seesaw mechanisms derived from (\ref{nm}) and (\ref{abcd}) as \be
M_{\mathrm{eff}}=M_L-M_D^TM_R^{-1}M_D=\left(%
\begin{array}{ccc}
  A & 0 & 0 \\
  0 & B_1 & C \\
  0 & C & B_2 \\
\end{array}%
\right), \label{Mef}\ee where \bea
A&=& a_L-\frac{a^2_D}{a_R},\crn
B_1&=&a_L-\frac{a_Rb^2_D-2a_Db_Db_R+a^2_D(a_R+d_R)}{a^2_R-b^2_R+a_Rd_R}\crn
B_2&=&B_1+d_L+\frac{2(b_Db_R-a_Da_R)d_D+(a^2_D-b^2_D)d_R-a_Rd^2_D}{a^2_R-b^2_R+a_Rd_R},\crn
C&=&b_L-\frac{(a^2_D+b^2_D)b_R-(2a_Da_R + a_Dd_R)b_D+(a_Db_R -a_Rb_D)d_D}{a^2_R-b^2_R+a_Rd_R}.\label{ABC}
\eea
We can diagonalize the mass matrix (\ref{Mef}) as follows $U^T_\nu
 M_{\mathrm{eff}} U_\nu=\mathrm{diag}(m_1,  m_2,  m_3)$, with
\bea m_1
&=&\fr 1 2 \left(B_1 + B_2 + \sqrt{(B_1 + B_2)^2+4C^2}\right),\crn
m_2&=&A,\label{m123}\\
m_3&=&\fr 1 2 \left(B_1 + B_2 - \sqrt{(B_1 + B_2)^2+4C^2}\right),\nn\eea
and the corresponding neutrino mixing matrix: \bea U_\nu=\left(%
\begin{array}{ccc}
  0 & 1 & 0 \\
  \fr{1}{\sqrt{K^2+1}} & 0 & \fr{K}{\sqrt{K^2+1}} \\
  -\fr{K}{\sqrt{K^2+1}} & 0 & \fr{1}{\sqrt{K^2+1}} \\
\end{array}%
\right).\left(%
\begin{array}{ccc}
  1 & 0 & 0 \\
 0 & 1 & 0 \\
 0 & 0& i \\
\end{array}%
\right),\label{Unu1}\eea
where
\bea
K&=&\frac{B_1 -B_2 -\sqrt{4 C^2 + (B_1-B_2)^2}}{2C}.\label{K}
\eea
Combining (\ref{Uclep}) and (\ref{Unu1}), we get the lepton mixing matrix:
\bea U^\dagger_L
U_\nu= \fr{1}{\sqrt{3}}\left(%
\begin{array}{ccc}
  \fr{1-K}{\sqrt{K^2+1}} & 1 &  \fr{1+K}{\sqrt{K^2+1}} \\
 \fr{\om(1-K\om)}{\sqrt{K^2+1}} & 1 &  \fr{\om(\om+K)}{\sqrt{K^2+1}} \\
   \fr{\om(\om-K)}{\sqrt{K^2+1}} & 1 &  \fr{\om(K\om+1)}{\sqrt{K^2+1}} \\
\end{array}%
\right).\left(%
\begin{array}{ccc}
  1 & 0 & 0 \\
 0 & 1 & 0 \\
 0 & 0& i \\
\end{array}%
\right).\label{Ulep}\eea \textbf{It is worth noting that in our
model, $K$ given in (\ref{K}) is an arbitrary number. Hence  in
general the lepton mixing matrix given in (\ref{Ulep}) is
different to $U_{HPS}$ in (\ref{Uhps}), but
 similar to the original version of trimaximal mixing considered
in \cite{TBM2} which is based on the $\Delta(27)$ group}.
\textbf{In the case where} $T_7$ is broken in $\mathrm{Identity}$
(Instead of $Z_3\rightarrow \mathrm{Identity}$) only by $s$, i.e,
without contribution of $\si'$ (or $\la'_\si=v'_\si=\La'_\si=0$),
the lepton mixing matrix (\ref{Ulep})  being equal to $U_{HPS}$ as
given in (\ref{Uhps}). This is a good features of $T_7$ with
tensor product $\underline{3} \otimes \underline{3}$ given in
(\ref{Tensorpr}).

In the standard Particle Data Group (PDG) parametrization, the lepton mixing
 matrix ($U_{PMNS}$) can be parametrized as
\be
       U_{PMNS} = \begin{pmatrix}
    c_{12} c_{13}     & -s_{12} c_{13}                    & - s_{13} e^{-i \delta}\\
    s_{12} c_{23}-c_{12} s_{23} s_{13}e^{i \delta} &
    c_{12} c_{23}-s_{12} s_{23} s_{13} e^{i \delta} & - s_{23} c_{13}\\
    s_{12} s_{23}+c_{12} c_{23} s_{13}e^{i \delta}&
    c_{12} s_{23}+s_{12} c_{23} s_{13} e^{i \delta}  &\,\,  c_{23} c_{13} \\
     \end{pmatrix} \times P. \label{Ulepg}
\ee where $P=\mathrm{diag}(1, e^{i \alpha}, e^{i \beta})$, and
$c_{ij}=\cos \theta_{ij}$, $s_{ij}=\sin \theta_{ij}$ with
$\theta_{12}$, $\theta_{23}$ and $\theta_{13}$ being the solar
angle, atmospheric angle and the reactor angle respectively.
$\delta$ is the Dirac CP violating phase while $\alpha$ and
$\beta$ are the two Majorana CP violating phases.

From the (\ref{Ulep}) and (\ref{Ulepg}) we rule out $\alpha=0,  \beta =\frac{\pi}{2}$, and the
lepton mixing matrix in (\ref{Ulep}) can be parameterized in three Euler's angles $\theta_{ij}$ as follows:
 \bea s_{13} e^{-i \delta}&=&\fr{-1-K}{\sqrt{3}\sqrt{K^2+1}},\label{s13}\\
 t_{12}&=&\frac{\sqrt{K^2+1}}{K-1},\label{t12}\\
  t_{23}&=&-\frac{\om+K}{1+K\om}.\label{t23}
\eea Substituting $\om =-\frac{1}{2}+i\frac{\sqrt{3}}{2}$ into
(\ref{t23}) yields: \bea K&=&k_1+ik_2,\crn
k_1&=&\frac{t^2_{23}-4t_{23}+1}{2(t^2_{23}-t_{23}+1)},\hs
k_2=\frac{\sqrt{3}(t^2_{23}-1)}{2(t^2_{23}-t_{23}+1)}.\label{K12}
\eea The expression (\ref{K12})  tells us that $k^2_1+k^2_2\equiv
|K|^2=1$. Combining (\ref{s13}) and (\ref{t12}) yields: \bea
e^{-i\delta}&=&\frac{1}{\sqrt{3}s_{13}t_{12}}\frac{1+K}{1-K}
=\frac{1}{\sqrt{3}s_{13}t_{12}}\left[\frac{1-k_1^2-k_2^2}{[(1-k_1)^2+k_2^2]}+
\frac{2k_2}{[(1-k_1)^2+k_2^2]}i\right]\crn
&=&-\frac{i(1-t_{23})}{s_{13}t_{12}(1+t_{23})}
\equiv\cos\delta-i\sin\delta \nn\eea or \bea \cos\delta&=&0,\hs
\sin\delta
=\frac{1-t_{23}}{s_{13}t_{12}(t_{23}+1)}.\label{sindel} \eea
Since $\cos\delta=0$ so that $\sin\delta$ must be equal to $\pm1$,
it is then $\delta =\frac{\pi}{2}$ or $\delta =\frac{3\pi}{2}$.
 Thus, our model predicts the maximal Dirac CP violating phase which
  is the same as \textbf{in Refs.} \cite{TBM2, MaximalCP}, and this is one of the most striking prediction
  of the model under consideration.

 \textbf{Up to now} the precise evaluation of $\theta_{23}$ is still an open
problem while $\theta_{12}$ and $\theta_{13}$ are now very
constrained \cite{PDG2012}. From (\ref{sindel}), our model can provide constraints on
$\theta_{23}$ from $\theta_{12}$ and $\theta_{13}$ which satisfy \cite{PDG2012} as follows.
\begin{itemize}
\item[(i)]  In the case $\delta = \frac{\pi}{2}$, from (\ref{sindel}) we have the relation
 among three Euler's angles as follows: \be
t_{23}=\frac{1-s_{13}t_{12}}{1+s_{13}t_{12}}.\label{relat1} \ee
In Fig. \ref{relat1v}, we have plotted the
values of $t_{23}$ as functions of
 $s_{13}$ and $t_{12}$ with $s_{13} \in (0.1585,
0.1590)$, $t_{12} \in (0.691, 0.692)$ . If  $s_{13}=0.1585\, (\theta_{13}=9.11^o)$ we
have the relation between $t_{23}$ and $t_{12}$ as shown in Fig. \ref{relat2v}.
\begin{figure}[h]
\begin{center}
\includegraphics[width=8.0cm, height=6.0cm]{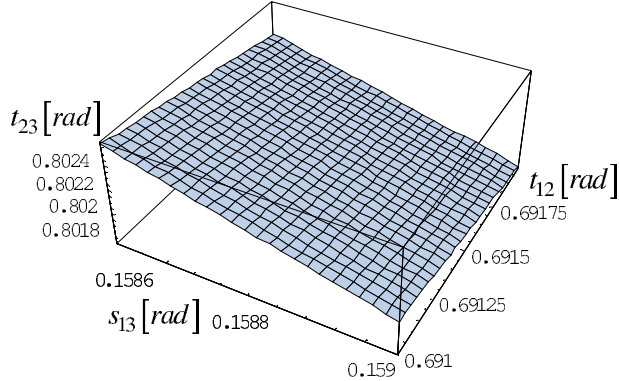}
\vspace*{-0.4cm} \caption[$t_{23}$ as functions of  $s_{13}$ and $t_{12}$. (a) $s_{13} \in (0.1585,
0.1590)$, $t_{12} \in (0.691, 0.692)$, (b) $s_{13} \in (0.1585,
0.1590)$, $t_{12} \in (1.447, 1.448)$]{$t_{23}$ as functions of  $s_{13}$ and $t_{12}$. (a) $s_{13} \in (0.1585,
0.1590)$, $t_{12} \in (0.691, 0.692)$, (b) $s_{13} \in (0.1585,
0.1590)$, $t_{12} \in (1.447, 1.448)$.}\label{relat1v}
\vspace*{-0.3cm}
\end{center}
\end{figure}

\begin{figure}[h]
\begin{center}
\includegraphics[width=8.0cm, height=5.5cm]{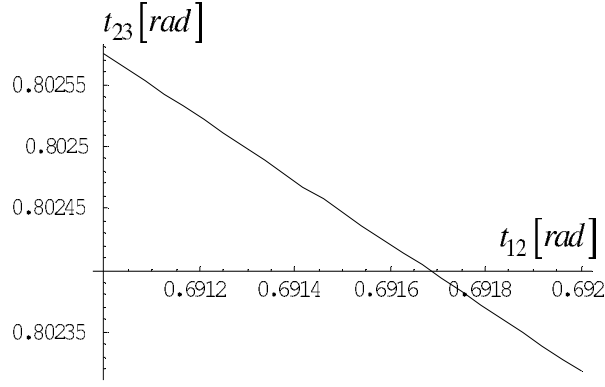}
\vspace*{-0.4cm} \caption[$t_{23}$ as a function of  $t_{12}$ with
$s_{13}=0.1585$ and $t_{12}\in (0.691, 0.692)$]{$t_{23}$ as a function
 of  $t_{12}$ with  $s_{13}=0.1585$ and $t_{12}\in (0.691, 0.692)$.}\label{relat2v}
\vspace*{-0.3cm}
\end{center}
\end{figure}
 For the best fit values of $\theta_{12}$ and $\theta_{13}$
 given in
 \cite{PDG2012}, $s_{13}=0.1585,\,\, t_{12}=0.691$ we obtain
 $t_{23}=0.802576 \, (\theta_{23}=38.75^o)$, and \bea
K=-0.930528 - 0.366221i, \hs (|K|=1). \label{Kvalues}\eea
The lepton mixing matrix in (\ref{Ulep}) then takes the form:
\be U\simeq\left(%
\begin{array}{ccc}
 0.831597 &\hs 0.57735 &\hs 0.157754 \\
-0.552417 &\hs 0.57735 &\hs -0.799061 \\
-0.27918 &\hs 0.57735 &\hs  0.641307 \\
\end{array}%
\right).\label{Ulepmix1}\ee These results also implies that in the
model under consideration, the value of the Jarlskog invariant
$J_{CP}$ which determines the magnitude of CP violation in
neutrino oscillations is determined \cite{Jarlskog}: \be
J_{CP}=\frac{1}{8}\cos\theta_{13}\sin2\theta_{12}\sin2\theta_{23}\sin2
\theta_{13}\sin\delta=0.03527.\label{J1}
\ee
\item[(ii)]
In the case $\delta =\frac{3\pi}{2}$, we have the relation among
three Euler's angles as follows: \be
t_{23}=\frac{1+s_{13}t_{12}}{1-s_{13}t_{12}}.\label{relat1} \ee
In Fig. \ref{relat3v}, we have plotted the
values of $t_{23}$ as a function of
 $s_{13}$ and $t_{12}$ with $s_{13} \in (0.1585,
0.1590)$, $t_{12} \in (0.691, 0.692)$.

\begin{figure}[h]
\begin{center}
\includegraphics[width=8.0cm, height=6.0cm]{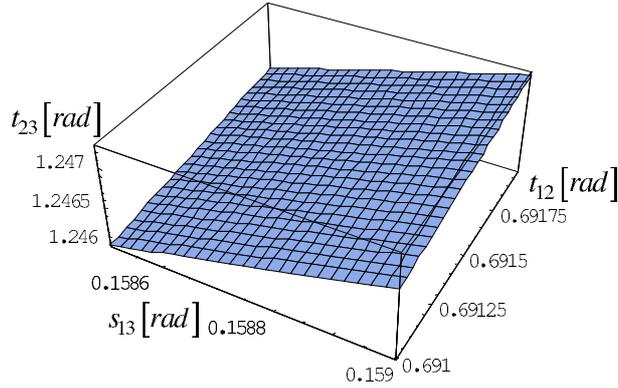}
\vspace*{-0.4cm} \caption[$t_{23}$ as a function of  $s_{13}$ and $t_{12}$ with $s_{13} \in (0.1585,
0.1590)$ and $t_{12} \in (0.691, 0.692)$]{$t_{23}$ as a function
 of  $s_{13}$ and $t_{12}$ with $s_{13} \in (0.1585,
0.1590)$ and $t_{12} \in (0.691, 0.692)$.}\label{relat3v}
\vspace*{-0.3cm}
\end{center}
\end{figure}
For the best fit values of $\theta_{12}$ and $\theta_{13}$
 given in \cite{PDG2012}, $s_{13}=0.1585,\,\, t_{12}=0.691$ we
 obtain $t_{23}=1.24599\, (\theta_{23}=51.25^o)$, and \be
K=-0.930527 + 0.366223i, \hs (|K|=1). \label{Kvalues1}\ee
The lepton mixing matrix in (\ref{Ulep}) in this case takes the form:
\be U\simeq\left(%
\begin{array}{ccc}
0.831597 &\hs 0.57735 &\hs-0.157755 \\
-0.279179 &\hs 0.57735 &\hs -0.641306 \\
-0.552418 &\hs 0.57735 &\hs 0.799061 \\
\end{array}%
\right),\label{Ulepmix1}\ee
and the value of the Jarlskog invariant $J_{CP}$ is determined \cite{Jarlskog}:
\be
J_{CP}=\frac{1}{8}\cos\theta_{13}\sin2\theta_{12}\sin2\theta_{23}
\sin2\theta_{13}\sin\delta\simeq-0.03527.\label{J1}
\ee
\end{itemize}

Until now values of neutrino masses (or the absolute neutrino
masses) as well as the mass ordering of neutrinos is unknown. The
tritium experiment \cite{Wein, Lobashev} provides an upper bound
on the absolute value of neutrino  mass
\bea
m_i\leq 2.2 \,
\mathrm{eV} \eea
 A more stringent bound was found from the analysis
of the latest cosmological data \cite{Tegmark} \be m_i\leq 0.6\,
\mathrm{eV},\label{upb} \ee
 while arguments from the growth of
large-scale structure in the early Universe yield the upper bound
\cite{Weiler} \be \sum^{3}_{i=1} m_i\leq 0.5\,  \mathrm{eV}.
\label{upbsum}\ee

The neutrino mass spectrum can be the normal mass hierarchy ($
|m_1|\simeq |m_2| < |m_3|$), the inverted hierarchy ($|m_3|< |m_1|\simeq |m_2|$)
 or nearly degenerate ($|m_1|\simeq |m_2|\simeq |m_3| $). The mass
ordering of neutrino depends on the sign of $\Delta m^2_{23}$
which is currently unknown. In the case of 3-neutrino mixing, in
the model under consideration,  the two possible signs of $\Delta
m^2_{23}$ correspond to two types of neutrino mass spectrum can be
provided. Combining (\ref{m123}) and the two experimental
constraints on squared mass differences of neutrinos
as shown in (\ref{PDG2012}) and the values of $K$ in (\ref{Kvalues})
or in (\ref{Kvalues1}), we have the solutions as shown bellows.

\subsection{Normal case ($\Delta m^2_{23}> 0$)}
\subsubsection{\label{sectionpi2}The case $\delta=\frac{\pi}{2}$}
In this case, combining (\ref{K}) and the values of $K$ in
(\ref{Kvalues}), we obtain \be B_1=B_2-(1.67146\times
10^{-7}+0.732442i)C. \label{B1B2v1} \ee Substituting $B_1$ from
(\ref{B1B2v1}) into (\ref{m123}) and combining with the two
experimental constraints on squared mass differences of neutrinos
as shown in (\ref{PDG2012}), \textbf{we get the solutions (in
[eV]) given in Appendix \ref{pi2}.}

From (\ref{B1B2v1}), (\ref{case1}) and (\ref{alphabeta}) we see that $A$ must be a real number in order to make
the  light neutrino masses $m_{1,2,3}$ to be real. In general, $B_{1,2}, C$ are complex numbers, $\al$ is
also a complex number but $\mathrm{Im}(\al) \ll \mathrm{Re} (\al)$; $m_{1,2}$ are real numbers, $m_3$ is
a complex number with $\mathrm{Im}(m_3) \ll \mathrm{Re} (m_3)$  but as will see below in the regions
of the model parameters $m_3$ is real number, too.

\begin{figure}[h]
\begin{center}
\includegraphics[width=12.0cm, height=5.0cm]{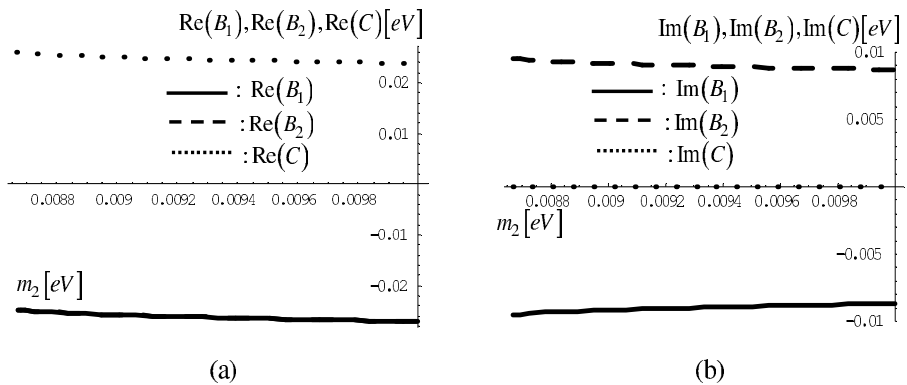}
\vspace*{-0.1cm} \caption[ (a) The real part of  $B_{1,2}$ and $C$ as functions of $m_2$, (b)
The imaginary part of  $B_{1,2}$ and $C$ as functions of $m_2$ in the case of $\Delta m^2_{23}> 0$]{ (a)
The real part of  $B_{1,2}$ and $C$ as functions of $m_2$, (b)
The imaginary part of  $B_{1,2}$ and $C$ as functions of $m_2$ in the case of $\Delta m^2_{23}> 0$.}\label{ReImmNcase1}
\end{center}
\end{figure}

\begin{figure}[h]
\bc
\includegraphics[width=12.0cm, height=10.0cm]{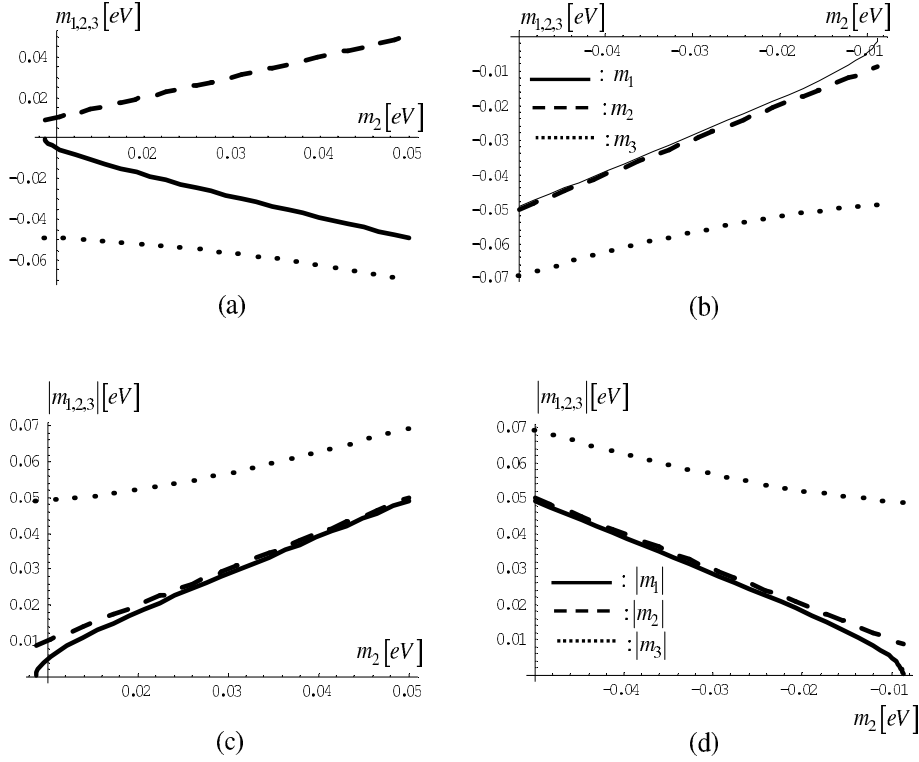}
\vspace*{-0.1cm} \caption[ The $m_{1,3}$ and $|m_{1,3}|$ as functions of $m_2$ in
 the case of $\Delta m^2_{23}> 0$.
(a) The $m_{1,3}$ as a function of $m_2$ with $m_2\in(0.00867, 0.05) \, \mathrm{eV}$,
(b) The $m_{1,3}$ as a function of $m_2$ with $m_2\in
(-0.05, -0.00867) \, \mathrm{eV}$; (c) The $|m_{1,3}|$ as a function of $m_2$ with
 $m_2\in(0.00867, 0.05) \, \mathrm{eV}$,  (d) The $|m_{1,3}|$ as a function of $m_2$
 with $m_2\in (-0.05, -0.00867) \, \mathrm{eV}$.]{\textbf{The $m_{1,3}$ and $|m_{1,3}|$ as functions of $m_2$ in
 the case of $\Delta m^2_{23}> 0$.
(a) The $m_{1,3}$ as a function of $m_2$ with $m_2\in(0.00867, 0.05) \, \mathrm{eV}$,
(b) The $m_{1,3}$ as a function of $m_2$ with $m_2\in
(-0.05, -0.00867) \, \mathrm{eV}$; (c) The $|m_{1,3}|$ as a function of $m_2$ with
 $m_2\in(0.00867, 0.05) \, \mathrm{eV}$,  (d) The $|m_{1,3}|$ as a function of $m_2$
 with $m_2\in (-0.05, -0.00867) \, \mathrm{eV}$.}}\label{m123N0case1}
\ec
\end{figure}

\begin{figure}[h]
\begin{center}
\includegraphics[width=12.0cm, height=5.5cm]{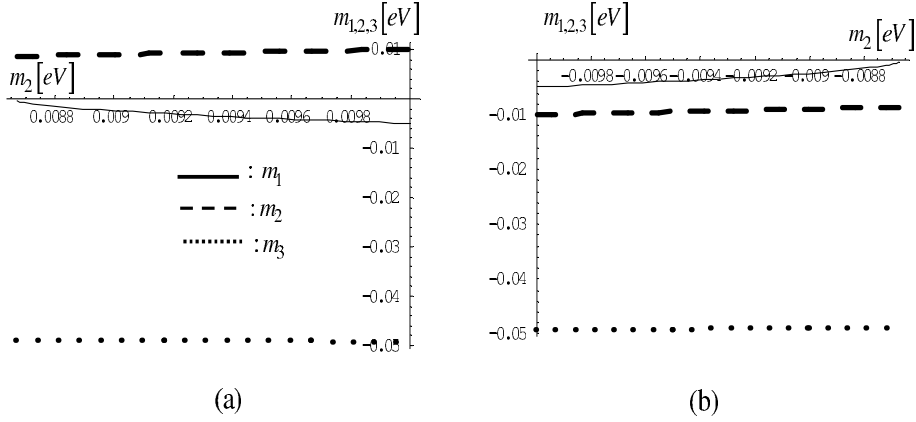}
\vspace*{-0.1cm} \caption[The $m_{1,3}$ and $|m_{1,3}|$ as functions of $m_2$ in the case
of $\Delta m^2_{23}> 0$.
(a) The $m_{1,3}$ as a function of $m_2$ with $m_2\in(0.00867, 0.01) \, \mathrm{eV}$,
(b) The $m_{1,3}$ as a function of $m_2$ with $m_2\in
(- 0.01, -0.00867) \, \mathrm{eV}$.]{\textbf{The $m_{1,3}$ and $|m_{1,3}|$ as functions of $m_2$ in the case
of $\Delta m^2_{23}> 0$.
(a) The $m_{1,3}$ as a function of $m_2$ with $m_2\in(0.00867, 0.01) \, \mathrm{eV}$,
(b) The $m_{1,3}$ as a function of $m_2$ with $m_2\in
(- 0.01, -0.00867) \, \mathrm{eV}$.}}\label{m123Ncase1}
\end{center}
\end{figure}

\begin{figure}[h]
\begin{center}
\includegraphics[width=12.0cm, height=5.0cm]{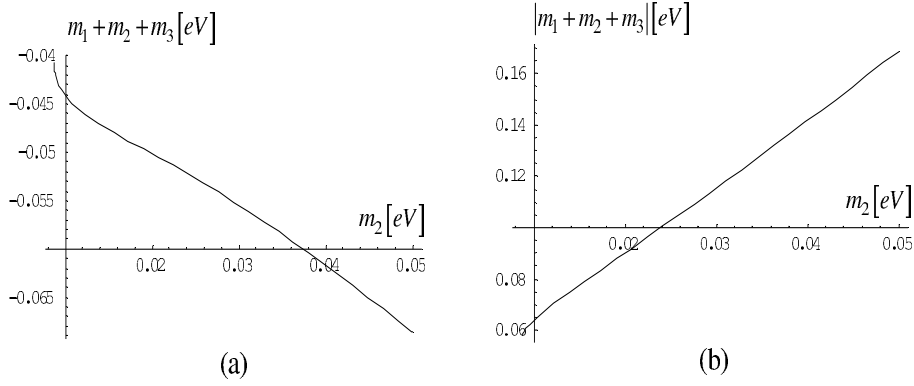}
\vspace*{-0.1cm} \caption[(a) The sum $\sum^3_{i=1}m_i$ as a function of
$A$ with $A \in (0.00867, 0.05)\,\mathrm{eV}$; (b) The sum $\sum^3_{i=1}|m_i|$ as a function of
$A$ with $A \in (0.00867, 0.05)\,\mathrm{eV}$ in the
case of $\Delta m^2_{23}> 0$]{(a) The sum $\sum^3_{i=1}m_i$ as a function of
$A$ with $A \in (0.00867, 0.05)\,\mathrm{eV}$; (b) The sum $\sum^3_{i=1}|m_i|$ as a function of
$A$ with $A \in (0.00867, 0.05)\,\mathrm{eV}$ in the
case of $\Delta m^2_{23}> 0$.}\label{m123Nscase1}
\end{center}
\end{figure}

The solutions in equations from (\ref{case1}) to (\ref{case4})
have the same absolute values of $m_{1,2, 3}$, the unique
difference is the sign of $m_{1,3}$. So, here we only consider in
detail the case in (\ref{case1}) \footnote{\textbf{The expressions
from (\ref{case1}) to (\ref{case4}) show that $m_{i}\hs (i=1,2,3)$
depends only on a parameter $A=m_2$ so we consider $m_{1,3}$ as
functions of $m_2$. However,  to have an explicit hierarchy on
neutrino masses, in the following figures, $m_2$ should be
included.}} Using the upper bound on the absolute value of
neutrino  mass (\ref{upb}) we can
 restrict the values of $A$:  $A \leq 0.6\,\mathrm{eV}$. However, in the case in (\ref{case1}),
  $A \in (0.0087, 0.01)\, \mathrm{eV}$ or $A \in (-0.01, -0.0087)\, \mathrm{eV}$ are good regions of $A$
that can reach the realistic neutrino mass hierarchy. In this region of $A$, $B_{1,2}$ and $C$
are complex numbers. The real parts and  the imaginary part of  $B_{1,2}$ and $C$ as functions
of $A$ (or $m_2$) are plotted in Figs. \ref{ReImmNcase1}a and \ref{ReImmNcase1}b, respectively.
In Figs. \ref{m123N0case1}a and \ref{m123N0case1}b, we have plotted the value $m_{1,3}$
as functions of $m_2$ with  $m_2 \in (0.00867, 0.05)\, \mathrm{eV}$ and $m_2
 \in (-0.05, -0.00867)\, \mathrm{eV}$. These figures shown
  that there exist allowed regions for values $m_2$ (or $A$) where either normal
  or quasi-degenerate neutrino masses spectrum achieved.
  The quasi-degenerate mass hierarchy obtained when $A$ lies
   in a region [$0.05\,\mathrm{eV} , +\infty$]  ($A$ increases
  but must be small enough because of the scale of $m_{1,2,3}$).
  As shown in Figs. \ref{m123Ncase1}a and \ref{m123Ncase1}b, the normal
  mass hierarchy will be obtained if $A$ takes the values around $(0.0087, 0.01)\,
   \mathrm{eV}$ or $(-0.01, -0.0087)\, \mathrm{eV}$.
The Figs. \ref{m123Nscase1}a and \ref{m123Nscase1}b give the sum
   $\sum^3_{i=1}m_i$ and $\sum^3_{i=1}|m_i|$ with $m_2 \in (0.0087, 0.05)\,\mathrm{eV}$, respectively.

\begin{figure}[h]
\begin{center}
\includegraphics[width=12.0cm, height=5.0cm]{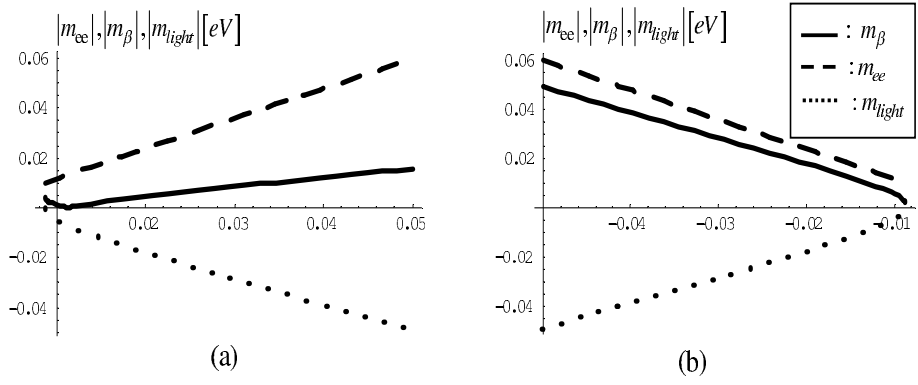}
\vspace*{-0.1cm} \caption[ The $|m_{ee}|$, $|m_{\beta}|$ and $|m_{light}|$ as functions of $m_2$
from (\ref{case1}) in the case of $\Delta m^2_{23}> 0$.
(a) $m_2\in(0.00867, 0.05) \,
\mathrm{eV}$,  (b) $m_2\in
(-0.05, -0.00867) \, \mathrm{eV}$.]{The $|m_{ee}|$, $|m_{\beta}|$ and $|m_{light}|$ as functions of $m_2$
from (\ref{case1}) in the case of $\Delta m^2_{23}> 0$.
(a) $m_2\in(0.00867, 0.05) \,
\mathrm{eV}$,  (b) $m_2\in
(-0.05, -0.00867) \, \mathrm{eV}$.}\label{mee.pi2.k--}
\end{center}
\end{figure}

\textbf{From the expressions (\ref{Ulep}) , (\ref{Kvalues1}) and (\ref{case1}), it is easily
 to obtain the effective mass $\langle m_{ee}\rangle$ governing neutrinoless double beta decay
 \cite{betdecay1, betdecay2,betdecay3,betdecay4,betdecay5, betdecay6},}
\bea
\langle m_{ee}\rangle &=& \mid\sum^3_{i=1} U_{ei}^2 m_i \mid \crn
&=&\frac{A}{3}-(0.333333-8.90616\times 10^{-18}i)\sqrt{4A^2-0.0003}\crn
&+&(0.0248863-1.26619\times 10^{-8}i)\sqrt{\Ga},\label{mee}\eea
\bea
\Ga&=&0.002245+(2 - 2.6469\times 10^{-23}i)A^2-(1.73176 + 1.22425\times 10^{-7}i)\sqrt{\ga},\\
\ga&=&(1.88579\times 10^{-7}+1.33377 i)(2.44973\times
10^{-19}+0.00866025i - Ai)\crn &\times& (4.79026\times
10^{-19}+0.0481664i+ A)(1.84151\times 10^{-18}-0.0481664i+ A)\crn
&\times&(0.00866025-6.69766\times 10^{-26}i + A).\label{Gaga} \eea
\be m_\beta = \sqrt{\sum^3_{i=1} |U_{ei}|^2 m_i^2 } \,, \label{mb}
\ee where \bea
\sum^3_{i=1} |U_{ei}|^2 m_i^2 &=&2.13672\times 10^{-6}+1.09955A^2 \label{mb}\\
        &-&(0.0430971 + 3.0467\times 10^{-9}i)\sqrt{\ga'}+0.0248863 \sqrt{4A^2-0.0003}\sqrt{\Ga'},\nn\eea
with \bea \ga'&=&-2.32077\times 10^{-7} + 3.28127\times
10^{-14}i+(0.00299432-  4.23359\times 10^{-10}i)A^2 \crn
&+&(1.33377 - 1.88579\times 10^{-7}i)A^4,\crn \Ga'&=&0.002245
+(2-2.64698\times 10^{-23}i)A^2-(1.73176+1.22425\times
10^{-7}i\sqrt{\ga'}). \label{Gagap} \eea \textbf{We also notice
that in the normal spectrum, $|m_1|\approx |m_2|<|m_3|$, so $m_1$
given in (\ref{case1}) is the lightest neutrino mass. Hence, it is
denoted as $m_{1}\equiv m_{light}$. In Figs. \ref{mee.pi2.k--}a
and \ref{mee.pi2.k--}b, we have plotted the value  $|m_{ee}|$,
$|m_{\beta}|$ and $|m_{light}|$ as functions of $m_2$ with $m_2
\in (0.0087, 0.05)\,\mathrm{eV}$ and $m_2 \in (-0.05,
-0.0087)\,\mathrm{eV}$, respectively.}

To get explicit values of the model
parameters, we assume $m_2=10^{-2}\, \mathrm{eV}$, which is safely small.
Then the other neutrino masses are explicitly
 given as $m_1\simeq-5.298\times 10^{-3}\, \mathrm{eV},\, m_2\simeq 10^{-2}\, \mathrm{eV},\,
  m_3\simeq -4.95 \times 10^{-2} \, \mathrm{eV}$ \textbf{and $|m_{ee}|\simeq 1.09
  \times 10^{-3} \, \mathrm{eV},\, |m_{\beta}|\simeq 1.178 \times 10^{-2} \, \mathrm{eV}$}.
  This solution means a normal
neutrino mass spectrum as mentioned above \textbf{and consistent with
the recent experimental data \cite{PDG2012, expbet1, expbet2}}. It follows that
\bea
C&\simeq&0.0237465-8.39362\times 10^{-10}i\simeq 0.0237465\,\mathrm{eV}, \crn
 B_1&=&-0.0270968 -0.00869645i, \hs
 B_2= -0.0276928 + 0.0232392i.\label{Vcase1}\eea

Furthermore, by assuming that
\be
\la_s=-\la_{\si}=-\la_{\si'}=-1\,\mathrm{eV},\, v_s=v_\si=-v'_\si,\, \La_s=-\La_\si=-\La'_\si=-v^2,\label{assum}\ee
 we obtain
 \bea
 A&=&2x, \hs C=y\left(-2+\frac{4x^2}{x^2-y^2+xz}\right),\crn
 B_1&=&x\left(2+\frac{4y^2}{x^2-y^2+xz}\right),\hs B_2=2(z-x)+\frac{4x^3}{x^2-y^2+xz}. \label{ABCcasse1}
 \eea
 Combining (\ref{Vcase1}) and (\ref{ABCcasse1}) yields:
$x\simeq5\times 10^{-3}$, $y\simeq (-4.52717-7.71265i)\times10^{-3}$,
\, $z\simeq (-10.4861 + 5.89481i)\times10^{-3}$.

\begin{figure}[h]
\begin{center}
\includegraphics[width=12.0cm, height=10.0cm]{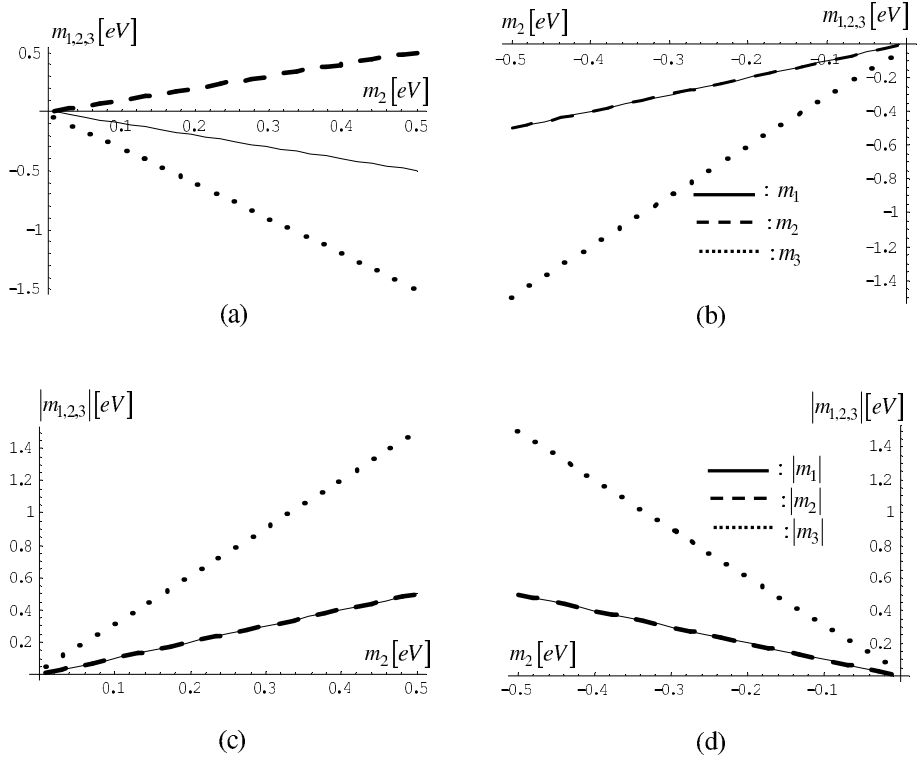}
\vspace*{-0.1cm} \caption[ The $m_{1,3}$ and $|m_{1,3}|$ as functions
 of $m_2$ from (\ref{case2}) in the case of $\Delta m^2_{23}> 0$.
(a) The $m_{1,3}$ as a function of $m_2$ with $m_2\in(0.00867, 0.5) \,
\mathrm{eV}$,  (b) The $m_{1,3}$ as a function of $m_2$ with $m_2\in
(-0.5, -0.00867) \, \mathrm{eV}$; (c) The $|m_{1,3}|$ as a function of $m_2$
with $m_2\in(0.00867, 0.5) \, \mathrm{eV}$,  (d) The $|m_{1,3}|$ as a
 function of $m_2$ with $m_2\in (-0.5, -0.00867) \, \mathrm{eV}$.]{The $m_{1,3}$
 and $|m_{1,3}|$ as functions of $m_2$ from (\ref{case2}) in the case of $\Delta m^2_{23}> 0$.
(a) The $m_{1,3}$ as a function of $m_2$ with $m_2\in(0.00867, 0.5) \, \mathrm{eV}$,
 (b) The $m_{1,3}$ as a function of $m_2$ with $m_2\in
(-0.5, -0.00867) \, \mathrm{eV}$; (c) The $|m_{1,3}|$ as a function of $m_2$ with $m_2
\in(0.00867, 0.5) \, \mathrm{eV}$,  (d) The $|m_{1,3}|$ as a function
of $m_2$ with $m_2\in (-0.5, -0.00867) \, \mathrm{eV}$.}\label{m123N0case2}
\end{center}
\end{figure}

Quite similar, the value $m_{1,3}$ and $|m_{1,3}|$ as functions of $m_2$ with $m_2
 \in (0.00867, 0.5)\, \mathrm{eV}$ and $m_2 \in (-0.5, -0.00867)\, \mathrm{eV}$ in the
 case in (\ref{case2}) was plotted in Figs. \ref{m123N0case2}a, \ref{m123N0case2}b,
 \ref{m123N0case2}c and \ref{m123N0case2}d. In this case, if $m_2=5\times 10^{-2}\,
 \mathrm{eV}$, which is safely small, the other neutrino masses are explicitly
 given as $m_1\simeq-4.925\times 10^{-2}\, \mathrm{eV},\, m_2\simeq 10^{-2}\,
 \mathrm{eV},\,  m_3\simeq -1,679 \times 10^{-1} \, \mathrm{eV}$. It follows that
  $C\simeq0.0637652-2.2539\times 10^{-9}i\simeq 0.0637652\,\mathrm{eV}, \,
 B_1\simeq-0.10858 - 0.0233521i, \, B_2\simeq -0.10858 + 0.0233521i$. Furthermore,
 with the assuming (\ref{assum}) we obtain $x\simeq2.5\times 10^{-2}$, $y\simeq
 (-1.244 - 4.172i)\times10^{-2}$, \, $z\simeq (-5.305+ 1.013i)\times10^{-2}$.

\begin{figure}[h]
\begin{center}
\includegraphics[width=12.0cm, height=10.0cm]{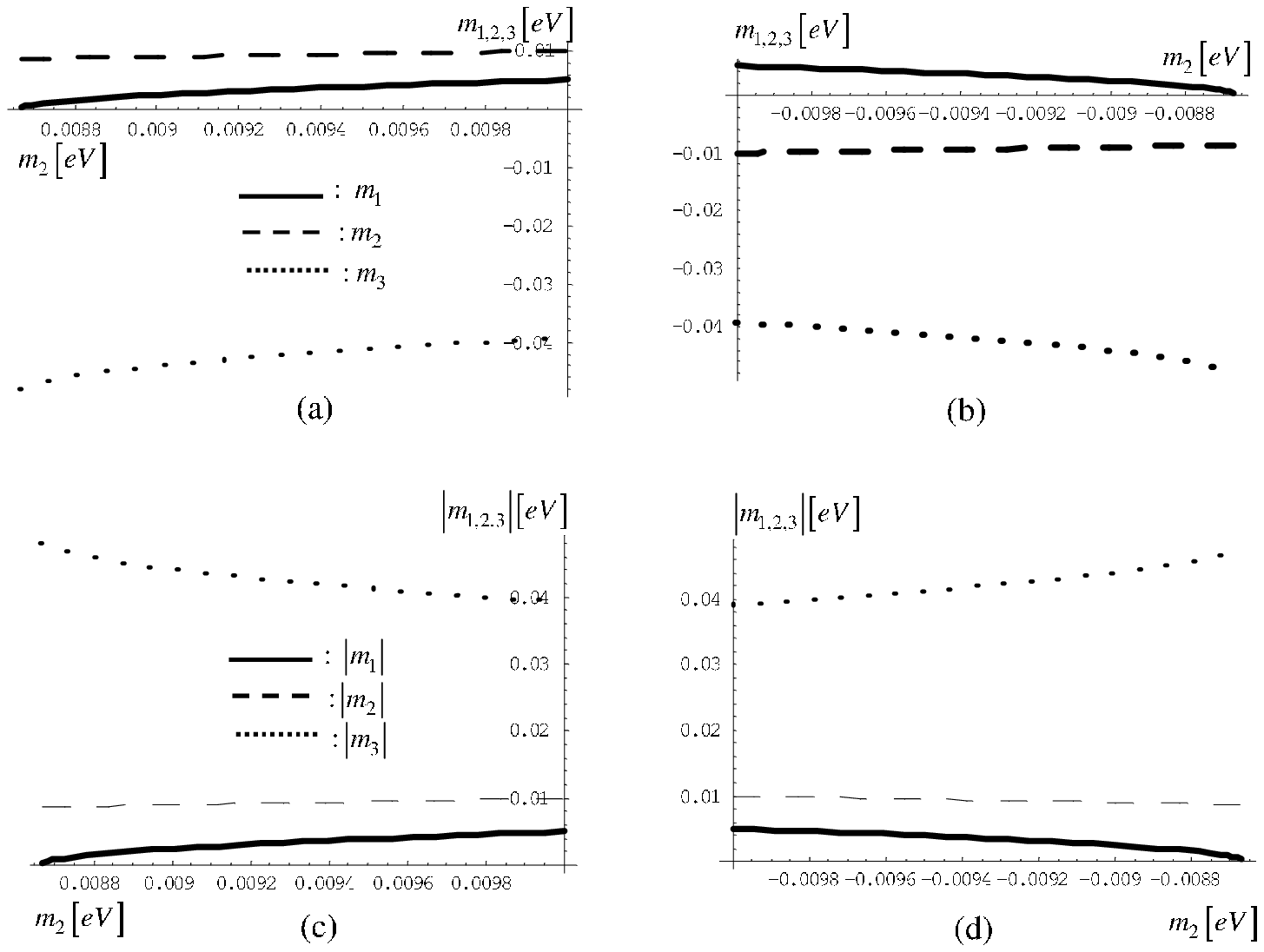}
\vspace*{-0.1cm} \caption[ The $m_{1,3}$ and $|m_{1,3}|$ as functions of $m_2$
from (\ref{case3}) in the case of $\Delta m^2_{23}> 0$.
(a) The $m_{1,3}$ as a function of $m_2$ with $m_2\in(0.00867, 0.01) \,
\mathrm{eV}$,  (b) The $m_{1,3}$ as a function of $m_2$ with $m_2\in
(-0.01, -0.00867) \, \mathrm{eV}$; (c) The $|m_{1,3}|$ as a function of $m_2$ with
$m_2\in(0.00867, 0.01) \, \mathrm{eV}$,  (d) The $|m_{1,3}|$ as a function of $m_2$
 with $m_2\in (-0.01, -0.00867) \, \mathrm{eV}$.]{The $m_{1,3}$ and $|m_{1,3}|$
 as functions of $m_2$ from (\ref{case3}) in the case of $\Delta m^2_{23}> 0$.
(a) The $m_{1,3}$ as a function of $m_2$ with $m_2\in(0.00867, 0.01) \,
\mathrm{eV}$,  (b) The $m_{1,3}$ as a function of $m_2$ with $m_2\in
(-0.5, -0.00867) \, \mathrm{eV}$; (c) The $|m_{1,3}|$ as a function of $m_2$ with
$m_2\in(0.00867, 0.01) \, \mathrm{eV}$,  (d) The $|m_{1,3}|$ as a function of
$m_2$ with $m_2\in (-0.01, -0.00867) \, \mathrm{eV}$.}\label{m123N0case3}
\end{center}
\end{figure}

In Figs. \ref{m123N0case3}a, \ref{m123N0case3}b, \ref{m123N0case3}c and \ref{m123N0case3}d
we have plotted the value $m_{1,3}$ and $|m_{1,3}|$ as functions of $m_2$ with
 $m_2 \in (0.00867, 0.01)\, \mathrm{eV}$ and $m_2 \in (-0.01, -0.00867)\, \mathrm{eV}$.

In this case, if we assume $m_2= 10^{-2}\, \mathrm{eV}$, which is safely small.
Then the other neutrino masses are explicitly
 given as $m_1\simeq5\times 10^{-5}\, \mathrm{eV},\hs m_3\simeq -3,92 \times
 10^{-2} \, \mathrm{eV}$. It follows that $C\simeq0.0237465- 8.39362\times
 10^{-10}i\simeq 0.0237465\,\mathrm{eV}, \,  B_1\simeq-0.0170968 -
 0.00869645i, \, B_2\simeq -0.0170967 + 0.00869645i$.
Furthermore, with the assuming (\ref{assum}) we obtain $x\simeq5\times
10^{-3}$, $y\simeq (-4.11978 - 5.98325i)\times10^{-3}$, \, $z\simeq (-8.36324 + 2.45329i)\times10^{-2}$.

\begin{figure}[h]
\begin{center}
\includegraphics[width=12.0cm, height=10.0cm]{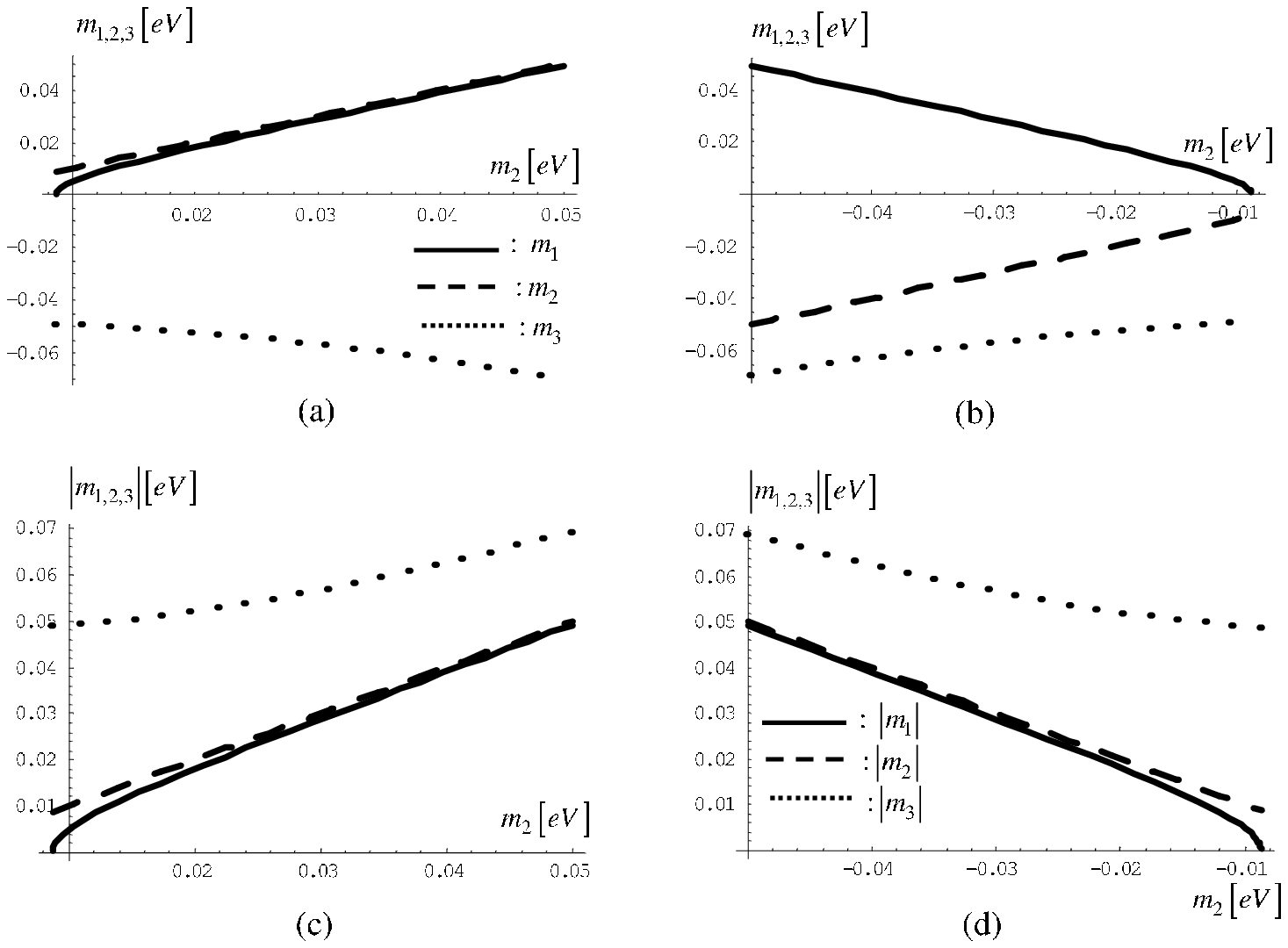}
\vspace*{-0.1cm} \caption[ The $m_{1,3}$ and $|m_{1,3}|$ as functions
of $m_2$ from (\ref{case4}) in the case of $\Delta m^2_{23}> 0$.
(a) The $m_{1,3}$ as a function of $m_2$ with $m_2\in(0.00867, 0.05) \,
\mathrm{eV}$,  (b) The $m_{1,3}$ as a function of $m_2$ with $m_2\in
(-0.05, -0.00867) \, \mathrm{eV}$; (c) The $|m_{1,3}|$ as a function of $m_2$ with
$m_2\in(0.00867, 0.05) \, \mathrm{eV}$,  (d) The $|m_{1,3}|$ as a function of
 $m_2$ with $m_2\in (-0.05, -0.00867) \, \mathrm{eV}$.]{ The $m_{1,3}$ and
 $|m_{1,3}|$ as functions of $m_2$ from (\ref{case4}) in the case of $\Delta m^2_{23}> 0$.
(a) The $m_{1,3}$ as a function of $m_2$ with $m_2\in(0.00867, 0.05) \,
\mathrm{eV}$,  (b) The $m_{1,3}$ as a function of $m_2$ with $m_2\in
(-0.05, -0.00867) \, \mathrm{eV}$; (c) The $|m_{1,3}|$ as a function of $m_2$ with
 $m_2\in(0.00867, 0.05) \, \mathrm{eV}$,  (d) The $|m_{1,3}|$ as a function of
 $m_2$ with $m_2\in (-0.05, -0.00867) \, \mathrm{eV}$.}\label{m123N0case4}
\end{center}
\end{figure}

In the case (\ref{case4}), the normal neutrino masses spectrum achieved with $A \in$ (0.00867, 0.05)
or $A\in (-0.05, -0.00867)$ as shown in Figs. \ref{m123N0case4}a,
\ref{m123N0case4}b, \ref{m123N0case4}c and \ref{m123N0case4}d.
In this case, if we assume $m_2= 10^{-2}\, \mathrm{eV}$, which is safely small.
Then the other neutrino masses are explicitly
 given as $m_1\simeq1.003\times 10^{-2}\, \mathrm{eV},\hs m_3\simeq 5.83
  \times 10^{-2} \, \mathrm{eV}$. It follows that
$C\simeq 0.0291198 -1.02929\times 10^{-9}i\simeq 0.0291198\,\mathrm{eV}, \,
 B_1\simeq-0.0320968 - 0.0106643i, \,
 B_2\simeq -0.0286423 + 0.0296257i$.
Furthermore, with the assuming (\ref{assum}), we obtain $x\simeq5\times 10^{-3}$,
$y\simeq (-5.48977 - 7.44641i)\times10^{-3}$, \, $z\simeq (-9.65158 + 8.47899i)\times10^{-2}$.

\subsubsection{The case $\delta=\frac{3\pi}{2}$}
In this case, combining (\ref{K}) and the values of $K$ in
(\ref{Kvalues1}), we obtain \be B_1=B_2+(2.01498\times
10^{-7}+0.732446i)C. \label{B1B2v2} \ee Substituting $B_1$ from
(\ref{B1B2v2}) into (\ref{m123}) and combining with the two
experimental constraints on squared mass differences of neutrinos
as shown in (\ref{PDG2012}), we \textbf{obtain four solutions (in
[eV])  given in Appendix \ref{3pi2}}.

\begin{figure}[h]
\begin{center}
\includegraphics[width=12.0cm, height=10.0cm]{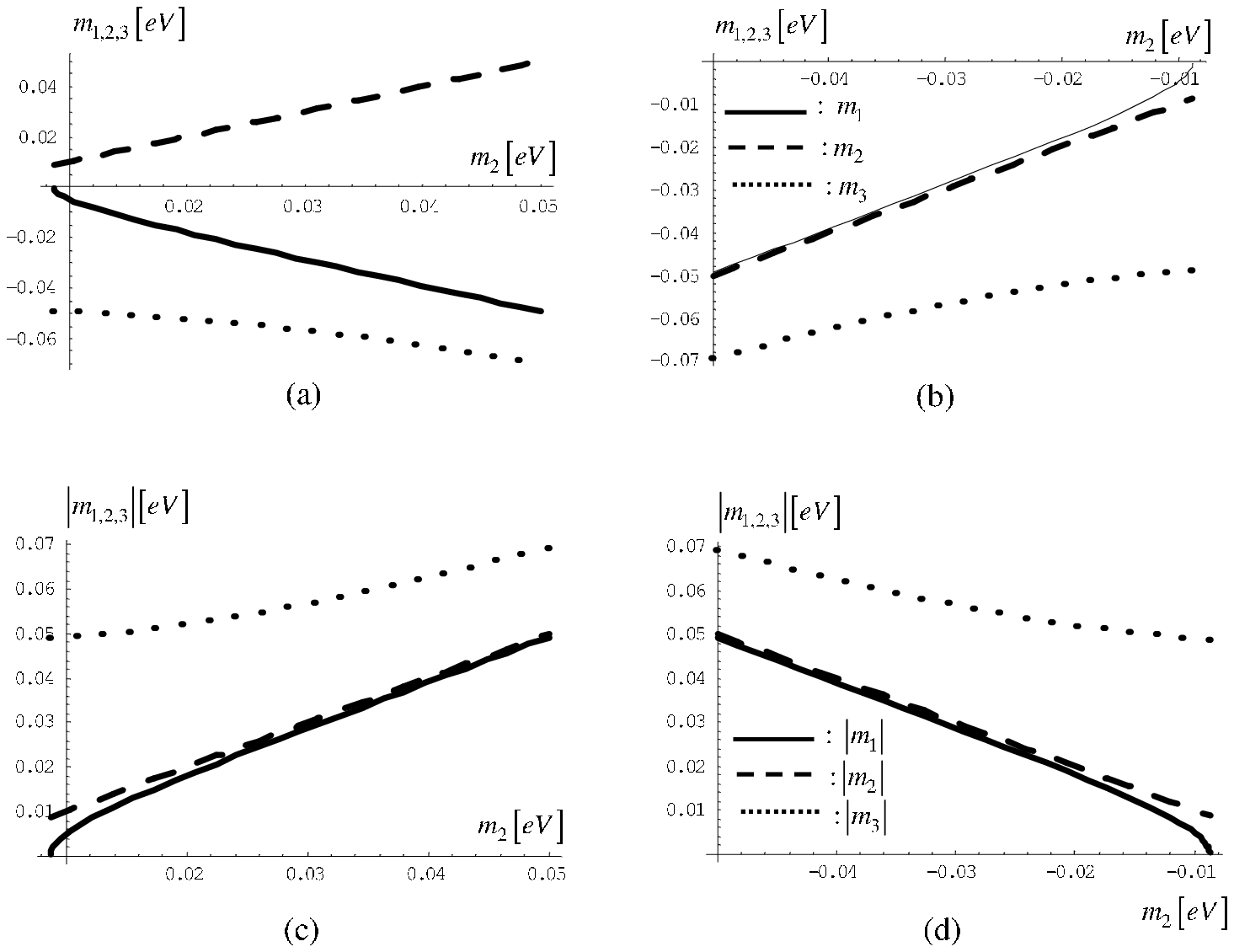}
\vspace*{-0.1cm} \caption[The $m_{1,3}$ and $|m_{1,3}|$ as functions of $m_2$
from (\ref{case5}) in the case of $\Delta m^2_{23}> 0$.
(a) The $m_{1,3}$ as a function of $m_2$ with $m_2\in(0.00867, 0.05) \,
\mathrm{eV}$,  (b) The $m_{1,3}$ as a function of $m_2$ with $m_2\in
(-0.05, -0.00867) \, \mathrm{eV}$; (c) The $|m_{1,3}|$ as a function of $m_2$ with
$m_2\in(0.00867, 0.05) \, \mathrm{eV}$,  (d) The $|m_{1,3}|$ as a function of
$m_2$ with $m_2\in (-0.05, -0.00867) \, \mathrm{eV}$.]{The $m_{1,3}$ and
$|m_{1,3}|$ as functions of $m_2$ from (\ref{case5}) in the case of $\Delta m^2_{23}> 0$.
(a) The $m_{1,3}$ as a function of $m_2$ with $m_2\in(0.00867, 0.05) \, \mathrm{eV}$,
(b) The $m_{1,3}$ as a function of $m_2$ with $m_2\in
(-0.05, -0.00867) \, \mathrm{eV}$; (c) The $|m_{1,3}|$ as a function of $m_2$ with
$m_2\in(0.00867, 0.05) \, \mathrm{eV}$,  (d) The $|m_{1,3}|$ as a function of
$m_2$ with $m_2\in (-0.05, -0.00867) \, \mathrm{eV}$.}\label{m123N0case5}
\end{center}
\end{figure}

Similar to the case $\delta=\frac{\pi}{2}$ in subsection \ref{sectionpi2}, in this case we also have four solutions
in which $m_{1, 3}$ have the same
absolute values, the unique difference is
the sign of $m_{1,3}$. So, we only consider one solution in the case (\ref{case5}).
 In this case, the value $m_{1,3}$ and $|m_{1,3}|$ as functions of $m_2$ Figs. \ref{m123N0case5}a,
 \ref{m123N0case5}b, \ref{m123N0case5}c, \ref{m123N0case5}d.
In this case, if we assume $m_2= 10^{-2}\, \mathrm{eV}$, which is safely small.
Then the other neutrino masses are explicitly
 given as $m_1\simeq8.997\times 10^{-3}\, \mathrm{eV},\hs m_3\simeq 5.00
  \times 10^{-3} \, \mathrm{eV}$. It follows that
$C\simeq 2.37465 -1.01188\times 10^{-7}i\,\mathrm{eV}, \,
 B_1\simeq-2.70967+0.869651i, \,
 B_2\simeq -2.76928 -2.32392i$.
Furthermore, with tha assuming (\ref{assum}), we obtain $x\simeq 5
\times 10^{-3}$, $y\simeq (-4.52717 + 7.71265i)\times10^{-3}$, \, $z
\simeq (-10.4861 - 5.8948i)\times10^{-2}$.

\subsection{Inverted case ($\Delta m^2_{23}< 0$)}
\begin{figure}[h]
\begin{center}
\includegraphics[width=14.0cm, height=5.5cm]{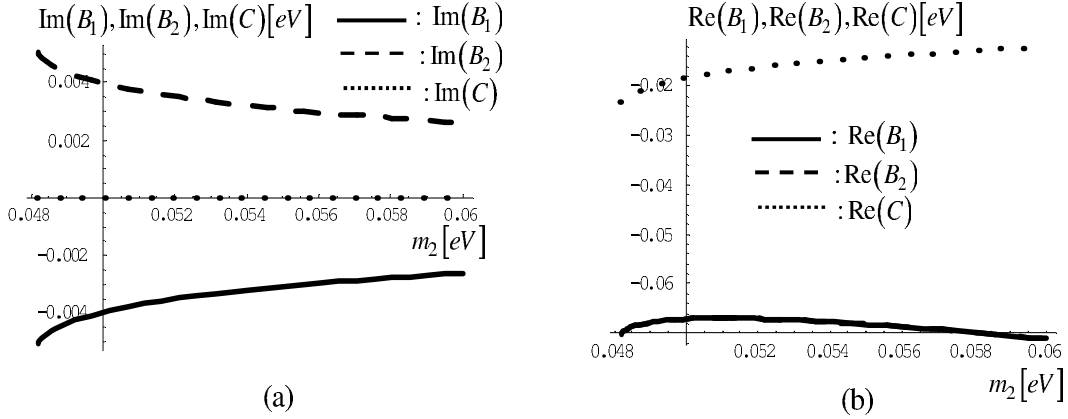}
\vspace*{-0.1cm} \caption[(a) The real part of  $B_{1,2}$ and $C$ as functions of $m_2$, (b)
The imaginary part of  $B_{1,2}$ and $C$ as functions of $m_2$ in the
case of $\Delta m^2_{23}< 0$ and $K=-0.930528-0.366221i$]{(a) The real
part of  $B_{1,2}$ and $C$ as functions of $m_2$, (b)
The imaginary part of  $B_{1,2}$ and $C$ as functions of $m_2$ in the
case of $\Delta m^2_{23}< 0$ and $K=-0.930528-0.366221i$.}\label{ImRemI}
\end{center}
\end{figure}

\begin{figure}[h]
\begin{center}
\includegraphics[width=12.0cm, height=10.0cm]{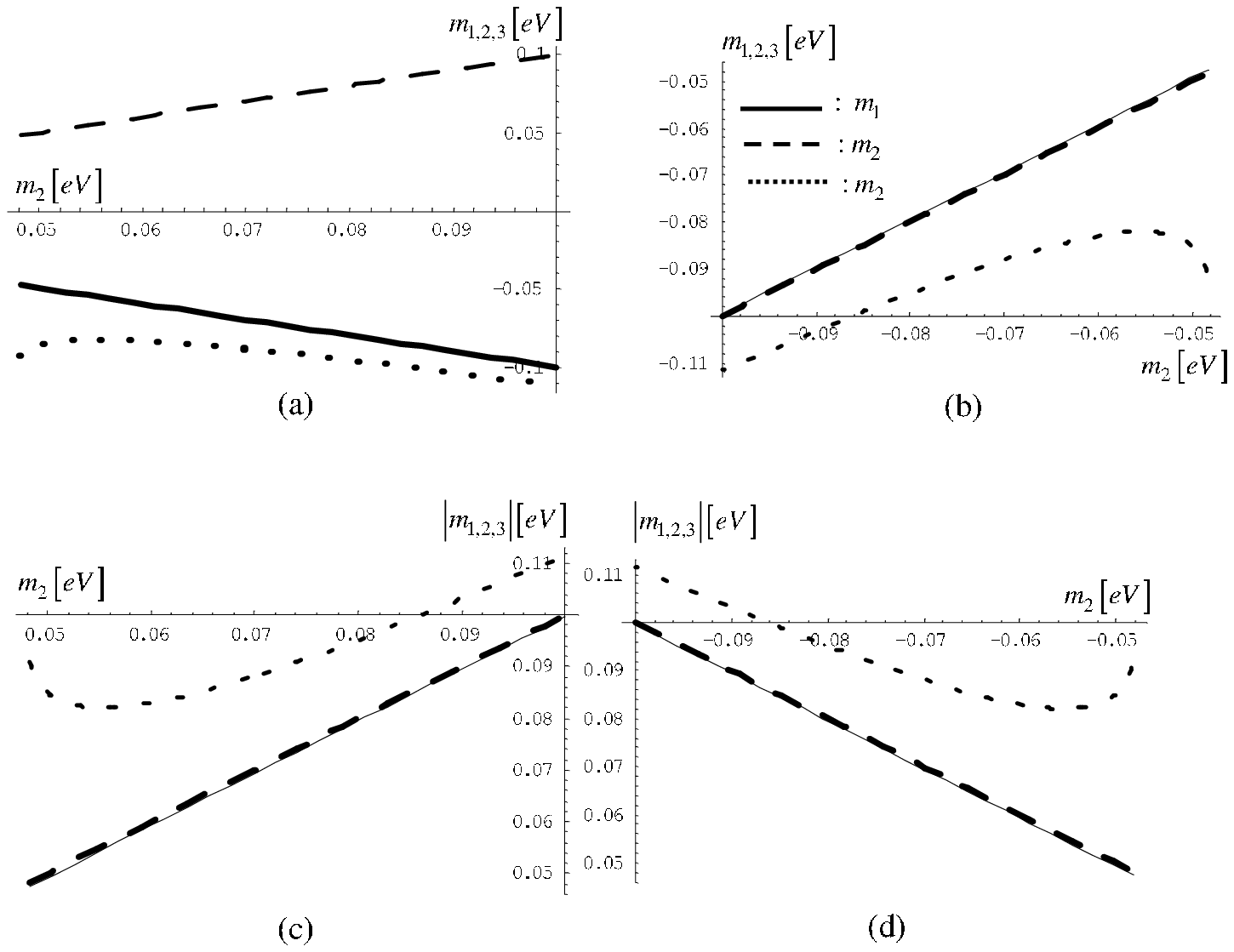}
\vspace*{-0.5cm} \caption[(a) The $m_{1,3}$ as functions of
$m_2$ with $m_2\in (0.0482, 0.1) \, \mathrm{eV}$ in the
case of $\Delta m^2_{23}< 0$ and $K=-0.930528-0.366221i$, (b) The $m_{1,3}$ as functions of
$m_2$ with $m_2\in (-0.1, -0.0482) \, \mathrm{eV}$ in the
case of $\Delta m^2_{23}< 0$ and $K=-0.930528-0.366221i$; (c) The $|m_{1,3}|$ as functions of
$m_2$ with $m_2\in (0.0482, 0.1) \, \mathrm{eV}$ in the
case of $\Delta m^2_{23}< 0$ and $K=-0.930528-0.366221i$, (d) The $|m_{1,3}|$ as functions of
$m_2$ with $m_2\in (-0.1, -0.0482) \, \mathrm{eV}$ in the
case of $\Delta m^2_{23}< 0$ and $K=-0.930528-0.366221i$]{(a) The $m_{1,3}$ as functions of
$m_2$ with $m_2\in (0.0482, 0.1) \, \mathrm{eV}$ in the
case of $\Delta m^2_{23}< 0$ and $K=-0.930528-0.366221i$, (b) The $m_{1,3}$ as functions of
$m_2$ with $m_2\in (-0.1, -0.0482) \, \mathrm{eV}$ in the
case of $\Delta m^2_{23}< 0$ and $K=-0.930528-0.366221i$; (c) The $|m_{1,3}|$ as functions of
$m_2$ with $m_2\in (0.0482, 0.1) \, \mathrm{eV}$ in the
case of $\Delta m^2_{23}< 0$ and $K=-0.930528-0.366221i$, (d) The $|m_{1,3}|$ as functions of
$m_2$ with $m_2\in (-0.1, -0.0482) \, \mathrm{eV}$ in the
case of $\Delta m^2_{23}< 0$ and $K=-0.930528-0.366221i$.}\label{m123I0}
\end{center}
\end{figure}

\begin{figure}[h]
\begin{center}
\includegraphics[width=12.0cm, height=10.0cm]{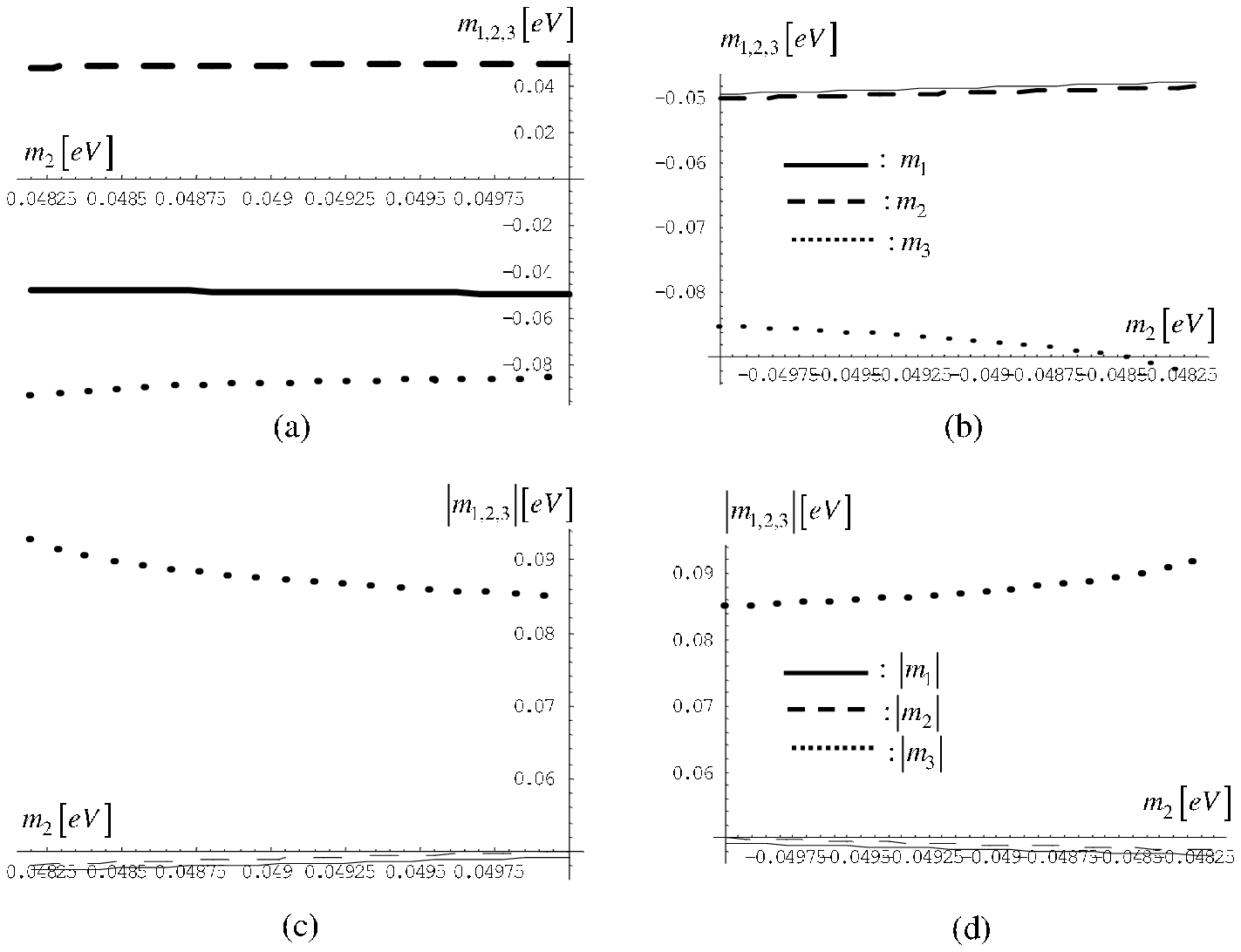}
\vspace*{-0.5cm} \caption[(a) The $m_{1,3}$ as functions of
$m_2$ with $m_2\in (0.0482, 0.05) \, \mathrm{eV}$ in the
case of $\Delta m^2_{23}< 0$ and $K=-0.930528-0.366221i$, (b) The $m_{1,3}$ as functions of
$m_2$ with $m_2\in (- 0.05,-0.0482) \, \mathrm{eV}$ in the
case of $\Delta m^2_{23}< 0$ and $K=-0.930528-0.366221i$; (c) The $|m_{1,3}|$ as functions of
$m_2$ with $m_2\in (0.0482, 0.05) \, \mathrm{eV}$ in the
case of $\Delta m^2_{23}< 0$ and $K=-0.930528-0.366221i$, (d) The $|m_{1,3}|$ as functions of
$m_2$ with $m_2\in (- 0.05,-0.0482) \, \mathrm{eV}$ in the
case of $\Delta m^2_{23}< 0$ and $K=-0.930528-0.366221i$.]{(a) The $m_{1,3}$ as functions of
$m_2$ with $m_2\in (0.0482, 0.05) \, \mathrm{eV}$ in the
case of $\Delta m^2_{23}< 0$ and $K=-0.930528-0.366221i$, (b) The $m_{1,3}$ as functions of
$m_2$ with $m_2\in (- 0.05,-0.0482) \, \mathrm{eV}$ in the
case of $\Delta m^2_{23}< 0$ and $K=-0.930528-0.366221i$; (c) The $|m_{1,3}|$ as functions of
$m_2$ with $m_2\in (0.0482, 0.05) \, \mathrm{eV}$ in the
case of $\Delta m^2_{23}< 0$ and $K=-0.930528-0.366221i$, (d) The $|m_{1,3}|$ as functions of
$m_2$ with $m_2\in (- 0.05,-0.0482) \, \mathrm{eV}$ in the
case of $\Delta m^2_{23}< 0$ and $K=-0.930528-0.366221i$.}\label{m123Icase1}
\end{center}
\end{figure}

\begin{figure}[h]
\begin{center}
\includegraphics[width=12.0cm, height=5.0cm]{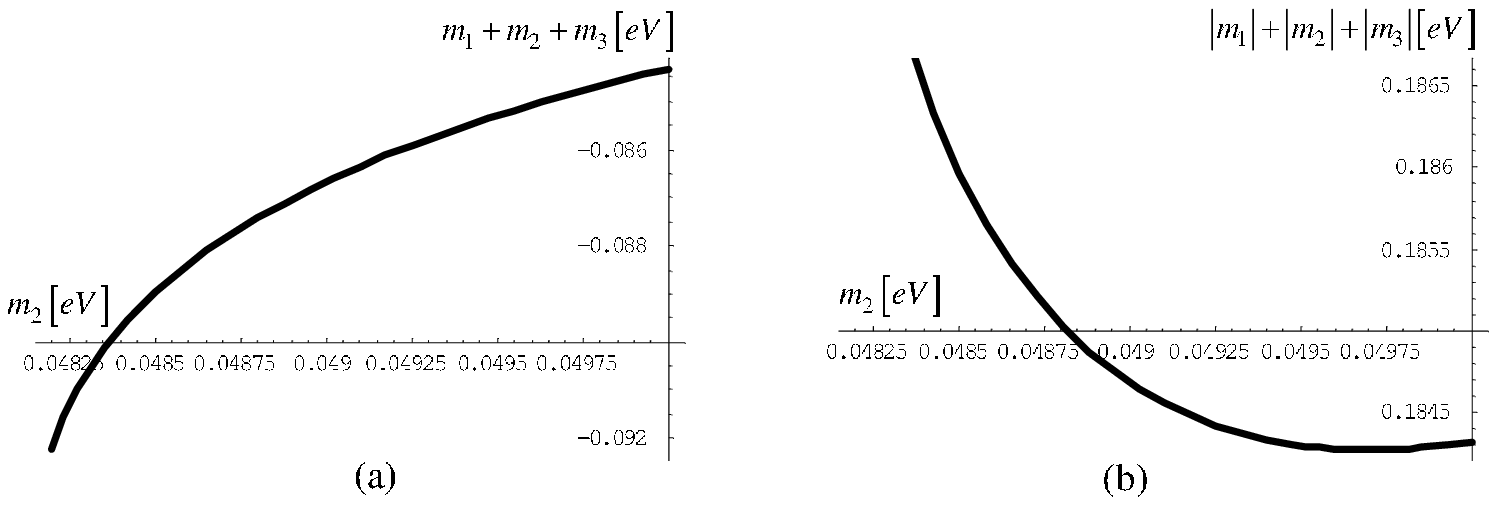}
\caption[(a) The sum $\sum^3_{i=1}m_i$ as a function of $m_2$ with $m_2\in (0.0482, 0.05)\,
\mathrm{eV}$ in the case of $\Delta m^2_{23}<
0$  and $K=-0.930528-0.366221i$, (b) The sum $\sum^3_{i=1}|m_i|$ as a function of
$m_2$ with $m_2\in (0.0482, 0.05)\,\mathrm{eV}$ in the case of $\Delta m^2_{23}<
0$  and $K=-0.930528-0.366221i$]{(a) The sum $\sum^3_{i=1}m_i$ as a function of $m_2$ with
 $m_2\in (0.0482, 0.05)\,\mathrm{eV}$ in the case of $\Delta m^2_{23}<
0$  and $K=-0.930528-0.366221i$, (b) The sum $\sum^3_{i=1}|m_i|$ as a function of
$m_2$ with $m_2\in (0.0482, 0.05)\,\mathrm{eV}$ in the case of $\Delta m^2_{23}<
0$  and $K=-0.930528-0.366221i$.}\label{m123Is}
\end{center}
\end{figure}

Similar to the normal case, in this case we also have four solutions in
which $m_{1,3}$ have the same
absolute values, the unique difference is
the sign of $m_{1,3}$. So, in the case $K=-0.930528-0.366221i$,
we only consider one solution in the form:
\bea
B_1&=&B_2-(1.67146\times 10^{-7}-0.732442i)C,\crn
C&=&0.5\sqrt{\al-2\sqrt{\beta}},\crn
B_2&=&-0.5\sqrt{4 A^2-0.0003}+(8.3573\times 10^{-8}+0.366221i)C\crn
&-&0.5\sqrt{(3.46353+2.4485\times 10^{-7})C^2},\crn
m_1&=&-0.5\sqrt{4 A^2-0.0003},\hs
m_2=A, \label{case1I}\\
m_3&=&-0.5\sqrt{4A^2-0.0003}\crn
&-&\sqrt{-0.002395+2.58494\times 10^{-26}i+(2-2.64698\times 10^{-23}i)
A^2-(1.73176+1.22425\times 10^{-7}i)\sqrt{\beta_1}}.\nn
\eea
where
\bea
\al_1&=&-0.00276597 + 1.95536\times 10^{-10}i+(2.30978-1.63287\times 10^{-7}i)A^2,\crn
\beta_1&=&(2.32077\times 10^{-7}-3.28127\times 10^{-14}i)-(0.00319439-4.51646\times 10^{-10}i)A^2\crn
&+&(1.33377-1.88579\times 10^{-7}i)A^4.\label{alphabeta1}
\eea
In Figs. \ref{ImRemI}a and \ref{ImRemI}b, we have plotted the real and the imaginary part of  $B_{1,2}$
and $C$ in (\ref{case1I}) as functions of $m_2$ with $m_2\in (0.0482, 0.05) \, \mathrm{eV}$, respectively.
 The $m_{1,3}$ and the absolute value $|m_{1,3}|$ as functions of $m_2$ with $m_2\in
(0.0482, 0.1) \, \mathrm{eV}$ are plotted in Figs. \ref{m123I0}a and \ref{m123I0}b, respectively.
These figures show that there exist
allowed regions for value of $m_2$ (or $A$) where
 either inverted or quasi-degenerate neutrino mass hierarchy achieved. The quasi-degenerate
  mass hierarchy obtained when $A$ lies in a region [$0.1\,\mathrm{eV} , +\infty$] 0r [$-\infty,
  -0.1\,\mathrm{eV}$] ($|A|$
 increases but must be small enough because of the scale of $m_{1,2,3}$). The
 inverted mass  hierarchy will be obtained if $A$ takes the values
  around $(0.0482, 0.05)\, \mathrm{eV}$ or $(- 0.05, -0.0482)\, \mathrm{eV}$ as shown
  in Figs. \ref{m123Icase1}a, \ref{m123Icase1}b, \ref{m123Icase1}c, \ref{m123Icase1}d.
The Figs. \ref{m123Is}a and \ref{m123Is}b give the sum $\sum^3_{i=1}m_i$ and $\sum^3_{i=1}|m_i|$ with
 $m_2 \in (0.0482, 0.05) \,\mathrm{eV}$, respectively.

 In similarity to the normal case, to get explicit values of the
model parameters, we assume $m_2=5\times10^{-2}\, \mathrm{eV}$,  which is safely small.
 Then the other neutrino masses are explicitly
 given as $ m_1 \simeq -4.925\times 10^{-2}\, \mathrm{eV}$ and $m_3\simeq8.507\times
 10^{-2} \, \mathrm{eV}$.
 It follows that $ C\simeq 0.0192514-6.80475\times 10^{-10}i \,\mathrm{eV},\, B_1\simeq
-0.0671582 - 0.00705026i\, \mathrm{eV},\, B_2\simeq-0.0671582 + 0.00705026i\, \mathrm{eV}.$
Furthermore,  with the assuming (\ref{assum}), we obtain $x\simeq 2.50\times 10^{-2}$,
$y\simeq (-0.365272-3.79824i)\times 10^{-2}$, $z\simeq (-3.41173 - 0.126634i) \times 10^{-2}$.

\section{Remark on breaking, VEVs and rho parameter}

 To obtain a realistic neutrino spectrum, in this work we
argue that both the breakings $T_7\rightarrow Z_3$ and
$T_7\rightarrow \{\mathrm{identity}\}$ must be taken place in neutrino sector
 while only the breaking $T_7\rightarrow Z_3$ is taken place in charged lepton and quark sectors.
  The quark masses at the tree-level can be fitted but then the CKM matrix is diagonal.

A breaking of
the lepton parity due to the odd VEVs $\langle \eta^0_3\rangle$, $\langle\chi^0_1\rangle$, or a
violation of $\mathcal{L}$ and/nor $S_3$ symmetry in terms of Yukawa interactions will disturb
the tree level matrix resulting in mixing between SM and
exotic quarks and/or possibly providing the desirable quark mixing
pattern  \cite{dlsvS4, dlnvS3, vlD4}. To get a realistic pattern of SM
 quarks mixing, we should add radiative correction
or use the effective six dimensional operators (for details, see Ref.  \cite{S3quark}).
However,  detailed study on this problem is out of the scope of this
work.

Note that $\La_\si, \La_s, \La_{\si'}$ are needed to the same
order and not to be so large that can naturally be taken at TeV
scale as the VEV $v_\chi$ of $\chi$. This is because $v_\si, v_s$
and $v_{\si'}$ carry lepton number, simultaneously breaking the
lepton parity which is naturally constrained to be much smaller
than the electroweak scale \cite{dlshA4, dlsvS4, dlnvS3,e331v1,
e331v2}. This is also behind a theoretical fact that $v_\chi$,
$\La_\si$ are scales for the gauge symmetry breaking in the first
stage from $\mathrm{SU}(3)_L\otimes \mathrm{U}(1)_X\rightarrow
\mathrm{SU}(2)_L\otimes \mathrm{U}(1)_Y$ in the original form of
3-3-1 models \cite{e331v1,e331v2, clong}. They will provide masses
for the new gauge bosons: $Z'$, $X$ and $Y$. Also, the exotic
quarks gain masses from $v_\chi$ while the neutral fermions masses
arise from $\La_\si, \La_s, \La_{\si'}$. The second stage of the
gauge symmetry breaking  from $\mathrm{SU}(2)_L\otimes
\mathrm{U}(1)_Y\rightarrow \mathrm{U}(1)_Q$ is achieved by the
electroweak scale VEVs such as $u, v$ responsible for ordinary
particle masses. In combination with those of type II seesaw as
determined, in our model, the following limit is often taken into
account  \cite{dlshA4, dlsvS4, dlnvS3,e331v1, e331v2}: \be
(\mathrm{eV})^2 \sim \la^2_\si,\la^2_s, \la^2_{\si'}\ll v^2_\si,
v^2_s, v^2_{\si'}\ll u^2, v^2\ll v_\chi^2\sim
\La^2_\si\sim\La^2_s\sim \La^2_{\si'}\sim(\mathrm{TeV})^2.
\label{limmitv}\ee

Our model contains a lot of $\mathrm{SU}(3)_L$ scalar triplets
that may modify the precision electroweak parameter such as $S, T,
U$ \cite{stu} and $\rho$ parameters. The most serious one can
result from the tree-level contributions to the $\rho$ parameter.
To see this let us approximate $W, Z$ mass and $\rho$ parameter
\footnote{We have used the notation $s_W=\sin\theta_W,
c_W=\cos\theta_W, t_W=\tan\theta_W$, and the continuation of the
gauge coupling constant $g$ of the $\mathrm{SU}(3)_L$ at the
spontaneous symmetry breaking point \cite{DongHLT, DongLongdh,
vlD4}, $t=\frac{3\sqrt{2}s_W}{\sqrt{3-4s^2_W}}$ was used.}: \bea
M^2_W &\simeq& \frac{g^2}{2}v^2_W,\hs M^2_Z \simeq \frac{g^2v^2_W}{2c^2_W},\crn
M^2_Y&\simeq&\frac{g^2}{2}\left(6\La^2_\si+2\La^2_s+2\La^2_{\si'}+v_\chi^2\right),
\label{MWXY}\eea and \bea \hs \rho=\fr{M^2_W}{c^2_W M^2_Z}\simeq
1+\fr{\lambda^2_s}{v^2_W}, \label{rho}\eea where $v^2_W\simeq
(3u^2+3v^2)= (246\, \textrm{GeV})^2$ is naturally given according
to (\ref{limmitv}) with $u\sim v\sim 100\,  \mathrm{GeV}$. Since
$\lambda_{s}=6v^2_\si+2v^2_s+2v^2_{\si'}$ is in eV scale
responsible for the observed neutrino masses, the $\rho$ in
(\ref{rho}) is absolutely close to the unity and in agreement with
the data \cite{PDG2012}.

The mixings between the charged gauge bosons $W-Y$ and the neutral ones $Z'-W_4$ are in the same
order since they are proportional to $\frac{v_\si}{\La_\si}$, and in the limit $\la_{s}, \la_{\si}, v_{s}, v_\si
\rightarrow 0$ these mixing angles tend to zero. In addition, from (\ref{limmitv}) and (\ref{MWXY}),
it follows that  $M^2_W$ is much smaller than $M^2_Y$.
\section{\label{conclus}Conclusions}
In this paper, we have constructed the $T_7$ model based on
$\mathrm{SU}(3)_C \otimes \mathrm{SU}(3)_L \otimes
\mathrm{U}(1)_X$ gauge symmetry responsible for fermion masses and
mixing. Neutrinos get masses from only anti-sextets which are in
triplets $\underline{3}$ and $\underline{3}^*$ under $T_7$. The
flavor mixing patterns and mass splitting are obtained without
perturbation. 
\textbf{The number of Higgs multiplets needed in order to
allow the fermions to gain masses are less than those of $S_3, S_4$ and $D_4$
\cite{dlnvS3, dlsvS4, vlD4}}. The tribimaximal form obtained
with the breaking $T_7 \rightarrow Z_3$ in charged lepton sector
and both $T_7 \rightarrow Z_3$ and $Z_3 \rightarrow
\{\mathrm{Identity}\}$ must be taken place in neutrino sector but
only apart in breakings $Z_3 \rightarrow \{\mathrm{Identity}\}$
(without contribution of $\si'$), and the upper bound on neutrino
mass  $\sum_{i=1}^3m_i$ at the level is presented. From the Dirac
CP violation phase we obtain the relation between Euler's angles which is consistent
with the experimental in PDG
2012. On the other hand, the realistic lepton mixing can be
obtained if both the direction for breakings $T_7 \rightarrow Z_3$
and $Z_3 \rightarrow \{\mathrm{Identity}\}$ are taken place in
neutrino sectors. The CKM  matrix is the identity matrix at the
tree-level. The Dirac CP violation phase $\delta$ is predicted to either
  $\frac{\pi}{2}$ or $\frac{3\pi}{2}$ which
 is maximal CP violation.

\section*{Acknowledgments}
This research is funded by Vietnam National Foundation for Science
and Technology Development (NAFOSTED) under grant number
103.01-2011.63.

\appendix

\section{\label{apa} $T_7$ group and Clebsch-Gordan coefficients}
The tetrahedral group $A_4$ has 12 elements and four equivalence classes with three
 inequivalent one-dimensional
 representations and one three-dimensional one, which is the smallest group with only a real
\underline{3} representation. The Frobenius group $T_7$ has 21 elements and five
 equivalence classes with three inequivalent one-dimensional
 representations and two three-dimensional once, which is
the smallest group with a pair of complex \underline{3} and $\underline{3}^*$
representations.  It is generated by
\bea
a &=& \left(%
\begin{array}{ccc}
\rho & 0 & 0 \\
0 & \rho^2 & 0 \\
 0 & 0 & \rho^4
 \end{array}%
\right),\hs
b = \left(%
\begin{array}{ccc}
0 & 1 & 0 \\
0 & 0 & 1 \\
 1 & 0 & 0 \end{array}%
\right),
\eea
where $\rho = \exp(2 \pi i /7)$, so that $a^7=1$, $b^3=1$, and $ab = ba^4$.
The character table of $T_7$ (with $\xi = -1/2 + i \sqrt{7}/2$) is given in table \ref{T7char}.

\begin{table}
\caption{\label{T7char}Character table of $T_7$ group}
\vspace{0.5cm}
\centerline{\begin{tabular}{|c|c|c|c|c|c|c|c|}
\hline
class & $n$ & $h$ & $\chi_1$ & $\chi_{1'}$ & $\chi_{1''}$ & $\chi_3$ &$\chi_{3^*}$ \\
\hline\hline
$C_1$ & 1 & 1 & 1 & 1 & 1 & 3 & 3 \\\hline
$C_2$ & 7 & 3 & 1 & $\omega$ & $\omega^2$ & 0 & 0 \\\hline
$C_3$ & 7 & 3 & 1 & $\omega^2$ & $\omega$ & 0 & 0 \\\hline
$C_4$ & 3 & 7 & 1 & 1 & 1 & $\xi$ & $\xi^*$ \\\hline
$C_5$ & 3 & 7 & 1 & 1 & 1 & $\xi^*$ & $\xi$ \\
\hline
\end{tabular}}
\end{table}

Let us put $\underline{3}(1,2,3)$ which means some
$\underline{3}$ multiplet such as $x=(x_1,x_2,x_3)\sim
\underline{3}$ or $y=(y_1,y_2,y_3)\sim \underline{3}$ and so on,
and similarly for the other representations. Moreover, the
numbered multiplets such as $(...,ij,...)$ mean $(...,x_i
y_j,...)$ where $x_i$ and $y_j$ are the multiplet components of
different representations $x$ and $y$, respectively. In the
following the components of representations in l.h.s will be
omitted and should be understood, but they always exist in order
in the components of decompositions in r.h.s. All the group multiplication
 rules of $T_7$ as given below.
\bea
\underline{1} \otimes \underline{1} &=&\underline{1}(11),\hs \underline{1}
\otimes \underline{1}' =\underline{1}'(11),
\hs\,\,\,\underline{1} \otimes \underline{1}'' =\underline{1}''(11), \crn
\underline{1}' \otimes \underline{1}''&=&\underline{1}(11), \hs \underline{1}' \otimes
\underline{1}'=\underline{1}''(11), \hs \underline{1}'' \otimes \underline{1}''=
\underline{1}'(11),\crn
 \underline{1} \otimes \underline{3}&=& \underline{3}(11,12,13), \,\,  \underline{1}' \otimes
 \underline{3}= \underline{3}(11,\om 12,\om^2 13), \,\,
\underline{1}'' \otimes \underline{3}= \underline{3}(11,\om^2 12,\om 13),\crn
 \underline{1} \otimes \underline{3}^*&=& \underline{3}^*(11,12,13), \,\,
 \underline{1}' \otimes \underline{3}^*= \underline{3}^*(11,\om 12,\om^2 13),\,\,
\underline{1}'' \otimes \underline{3}^*= \underline{3}^*(11,\om^2 12,\om 13), \crn
\underline{3} \otimes \underline{3}&=&\underline{3} (33,11,22)\oplus
\underline{3}^* (23,31,12) \oplus
\underline{3}^* (32,13,21) , \label{Tensorpr}\\
\underline{3}^* \otimes \underline{3}^* &=&\underline{3}^* (33,11,22)\oplus
 \underline{3}(23,31,12) \oplus
\underline{3}(32,13,21),\crn
\underline{3} \otimes \underline{3}^* &=&\underline{1} (1 1 + 2 2 + 3 3)\oplus
\underline{1}' (1 1 + \omega 2 2 + \omega^2 3 3) \crn
&\oplus& \underline{1}'' (1 1 + \omega^2 2 2 + \omega 3 3) \oplus \underline{3} (2 1, 3 2, 1 3)
\oplus \underline{3}^* (1 2, 2 3, 3 1) .\nn
\eea
Note that $\underline{3} \times \underline{3} \times \underline{3}$ has two
invariants and $\underline{3} \times \underline{3} \times \underline{3}^*$
has one invariant.

\section{\label{apb}The numbers}
In the following we will explicitly point out the lepton number
($L$) and lepton parity ($P_l$) of the model particles (notice
that the family indices are suppressed): \bc
\begin{tabular}{|c|c|c|}
  \hline
Particles & $L$ & $P_l$  \\ \hline
 $N_R$, $u$, $d$,  $\phi^+_1$,$\phi'^+_1$, $\phi^0_2$,$\phi'^0_2$,
  $\eta^0_1$,$\eta'^0_1$, $\eta^-_2$,$\eta'^-_2$
  $\chi^0_3$, $\sigma^0_{33}$, $s^0_{33}$ & 0 & 1  \\ \hline
  $\nu_L$, $l$, $U$, $D^*$, $\phi^+_3$,$\phi'^+_3$, $\eta^0_3$,
  $\eta'^0_3$, $\chi^{0*}_1$, $\chi^+_2$,
   $\sigma^0_{13}$,
   $\sigma^+_{23}$, $s^0_{13}$, $s^+_{23}$ & $-1$ & $-1$
   \\ \hline
   $\sigma^{0}_{11}$, $\sigma^{+}_{12}$, $\sigma^{++}_{22}$,
   $s^{0}_{11}$, $s^{+}_{12}$, $s^{++}_{22}$ & $-2$ & 1 \\ \hline
\end{tabular}\ec
\section{\label{pi2}The solutions with $\delta=\frac{\pi}{2}$ in the normal case}
 \bit \item The first case:
 \bea
C&=&0.5\sqrt{\al-2\sqrt{\beta}},\crn
B_2&=&-0.5\sqrt{4 A^2-0.0003}+(8.3573\times 10^{-8}+0.366221i)C\crn
&-&0.5\sqrt{(3.46353+2.4485\times 10^{-7})C^2},\crn
m_1&=&-0.5\sqrt{4 A^2-0.0003},\hs
m_2=A, \label{case1}\\
m_3&=&-0.5\sqrt{4 A^2-0.0003}\crn
&-&\sqrt{0.002245+(2-2.64698\times 10^{-23}i)A^2-(1.73176+1.22425\times 10^{-7}i)\sqrt{\beta}}.\nn
\eea
\item The second case:
  \bea
C&=&0.5\sqrt{\al+2\sqrt{\beta}},\crn
B_2&=&-0.5\sqrt{4 A^2-0.0003}+(8.3573\times 10^{-8}+0.366221i)C\crn
&-&0.5\sqrt{(3.46353+2.4485\times 10^{-7})C^2},\crn
m_1&=&-0.5\sqrt{4 A^2-0.0003},\hs
m_2=A, \label{case2}\\
m_3&=&-0.5\sqrt{4 A^2-0.0003}\crn
&-&\sqrt{0.002245+(2-2.64698\times 10^{-23}i)A^2+(1.73176+1.22425\times 10^{-7}i)\sqrt{\beta}},\nn
\eea
\item The third case:
 \bea
C&=&-0.5\sqrt{\al-2\sqrt{\beta}},\crn
B_2&=&0.5\sqrt{4 A^2-0.0003}+(8.3573\times 10^{-8}+0.366221i)C\crn
&-&0.5\sqrt{(3.46353+2.4485\times 10^{-7})C^2},\crn
m_1&=&0.5\sqrt{4 A^2-0.0003},\hs
m_2=A, \label{case3}\\
m_3&=&0.5\sqrt{4 A^2-0.0003}\crn
&-&\sqrt{0.002245+(2-2.64698\times 10^{-23}i)A^2+(1.73176+1.22425\times 10^{-7}i)\sqrt{\beta}},\nn
\eea
\item The fourth case:
 \bea
C&=&0.5\sqrt{\al+2\sqrt{\beta}},\crn
B_2&=&0.5\sqrt{4 A^2-0.0003}+(8.3573\times 10^{-8}+0.366221i)C\crn
&-&0.5\sqrt{(3.46353+2.4485\times 10^{-7})C^2},\crn
m_1&=&0.5\sqrt{4 A^2-0.0003},\hs
m_2=A, \label{case4}\\
m_3&=&0.5\sqrt{4 A^2-0.0003}\crn
&-&\sqrt{0.002245+(2-2.64698\times 10^{-23}i)A^2+(1.73176+1.22425\times 10^{-7}i)\sqrt{\beta}},\nn
\eea
where
\bea
\al &=&(0.00259273-1.8329\times 10^{-10}i) + (2.30978-1.63287\times 10^{-7}i)A^2,\crn
\beta&=&-2.32077\times 10^{-7}+3.28127\times 10^{-14}i \label{alphabeta}\\
&+&(0.00299432-4.23359\times 10^{-10}i)A^2+(1.33377-1.88579\times 10^{-7}i)A^4.\nn
\eea
\eit
\section{\label{3pi2}The solutions with $\delta=\frac{3\pi}{2}$ in the normal case}
 \begin{itemize}
 \item The first case:
 \bea
C&=&0.5\sqrt{\al'-2\sqrt{\beta'}},\crn
B_2&=&-0.5\sqrt{4 A^2-0.0003}+(1.00749\times 10^{-7}+0.366223i)C\crn
&-&0.5\sqrt{(3.46352+2.95173\times 10^{-7})C^2},\crn
m_1&=&-0.5\sqrt{4 A^2-0.0003},\hs
m_2=A, \label{case5}\\
m_3&=&-0.5\sqrt{4 A^2-0.0003}\crn
&-&\sqrt{0.002245+2.58494\times 10^{-26}i+2A^2-(1.73176+1.47587\times 10^{-7}i)\sqrt{\beta'}}.\nn
\eea

\item The second case:
  \bea
C&=&0.5\sqrt{\al+2\sqrt{\beta}},\crn
B_2&=&-0.5\sqrt{4 A^2-0.0003}+(1.00749\times 10^{-7}+0.366223i)C\crn
&-&0.5\sqrt{(3.46352+2.95173\times 10^{-7})C^2},\crn
m_1&=&-0.5\sqrt{4 A^2-0.0003},\hs
m_2=A, \label{case6}\\
m_3&=&-0.5\sqrt{4 A^2-0.0003}\crn
&-&\sqrt{0.002245+2.58494\times 10^{-26}i+2A^2+(1.73176+1.47587\times 10^{-7}i)\sqrt{\beta'}}.\nn
\eea
\item The third case:
 \bea
C&=&0.5\sqrt{\al'-2\sqrt{\beta'}},\crn
B_2&=&0.5\sqrt{4 A^2-0.0003}-(1.00749\times 10^{-7}+0.366223i)C\crn
&-&0.5\sqrt{(3.46352+2.95173\times 10^{-7})C^2},\crn
m_1&=&0.5\sqrt{4 A^2-0.0003},\hs
m_2=A, \label{case7}\\
m_3&=&0.5\sqrt{4 A^2-0.0003}\crn
&-&\sqrt{0.002245+2.58494\times 10^{-26}i+2A^2-(1.73176+1.47587\times 10^{-7}i)\sqrt{\beta'}},\nn
\eea
\item The fourth case:
 \bea
C&=&0.5\sqrt{\al'+2\sqrt{\beta'}},\crn
B_2&=&0.5\sqrt{4 A^2-0.0003}-(1.00749\times 10^{-7}+0.366223i)C\crn
&-&0.5\sqrt{(3.46352+2.95173\times 10^{-7})C^2},\crn
m_1&=&0.5\sqrt{4 A^2-0.0003},\hs
m_2=A, \label{case8}\\
m_3&=&0.5\sqrt{4 A^2-0.0003}\crn
&-&\sqrt{0.002245+(2-2.64698\times 10^{-23}i)A^2+(1.73176+1.47587\times 10^{-7}i)\sqrt{\beta'}},\nn
\eea
where
\bea
\al' &=&(0.00259274-2.20962\times 10^{-10}i) + (2.30979-1.96848\times 10^{-7}i)A^2,\crn
\beta'&=&-2.32078\times 10^{-7}+3.95569\times 10^{-14}i \label{alphabetap}\\
&+&(0.00299433-5.10375\times 10^{-10}i)A^2+(1.33378-2.27338\times 10^{-7}i)A^4.\nn
\eea
 \end{itemize}


\begin{thebibliography}{99}
\bibitem{altar1}  G. Altarelli, \emph{An Overview of Neutrino Mixing}, ArXiv: 1210.3467, Nucl. Phys. B
Proceedings Supplement (2012).

\bibitem{altar2}  G. Altarelli,{\it Neutrino Mixing: Theoretical Overview}, ArXiv: 1304.5047
(2013), and references therein.



\bibitem{hps1} P. F. Harrison, D.  H. Perkins and W. G. Scott, Phys. Lett. B {\bf 530}, 167 (2002).
\bibitem{hps2} Z. Z. Xing, Phys. Lett. B {\bf 533}, 85 (2002).
\bibitem{hps3} X. G. He and A. Zee, Phys. Lett. B {\bf 560}, 87 (2003).
\bibitem{hps4} X. G. He and A. Zee, Phys. Rev. D {\bf 68}, 037302 (2003).


\bibitem{smirnov} A. Yu. Smirnov, \emph{Neutrino 2012: Outlook - theory},
ArXiv: 1210.4061,  Nucl. Phys. B Proceedings Supplement (2012).


\bibitem{PDG2012} J. Beringer,  \emph{et al.}(2012),
    \emph{Review of Particle Physics} (Particle Data Group), Phys. Rev. D. 86, 010001.
\bibitem{PDG1} T. Schwetz, M. Tortola, and J. Valle (2011), New J. Phys. 13,
109401, arXiv: 1108.1376 [hep-ph].
\bibitem{PDG2} K. Abe {\it et al.}(2011) [T2K Collaboration], Phys. Rev. Lett. 107, 041801.
\bibitem{PDG3} P. Adamson {\it et al.} (2011) [MINOS Collaboration], Phys. Rev.Lett. 107,
181802, arXiv: 1108.0015 [hep-ex].
\bibitem{PDG4}  Fogli, G.L. et al. (2011),  Phys. Rev. D84, 053007,  arXiv: 1106.6028 [hep-ph]



\bibitem{CKM} N. Cabibbo, Phys Rev. Lett. 10, 531 (1963).
\bibitem{CKM1} M. Kobayashi and T. Maskawa, Prog. Theor. Phys. 49, 652 (1973).



\bibitem{A41} E. Ma and G. Rajasekaran, Phys. Rev. D {\bf 64}, 113012 (2001).
\bibitem{A42} K. S. Babu, E. Ma and J. W. F. Valle, Phys. Lett. B \textbf{552}, 207 (2003).
\bibitem{A43} G. Altarelli and F. Feruglio, Nucl. Phys. B \textbf{720}, 64 (2005).
\bibitem{A44} E. Ma, Phys. Rev. D {\bf 73}, 057304 (2006).
\bibitem{A45} X. G. He, Y. Y. Keum and R. R. Volkas, JHEP {\bf 0604}, 039 (2006).\
\bibitem{A46} S. Morisi, M. Picariello, and E. Torrente-Lujan, Phys. Rev. D {\bf 75}, 075015 (2007).
\bibitem{A47} C. S. Lam, Phys. Lett. B {\bf 656}, 193 (2007).
\bibitem{A48} F. Bazzocchi, S. Kaneko and S. Morisi, JHEP \textbf{0803}, 063 (2008).
\bibitem{A49} A. Blum, C. Hagedorn, and M. Lindner, Phys. Rev. D {\bf 77}, 076004 (2008).
\bibitem{A410} F. Bazzochi, M. Frigerio, and S. Morisi, Phys. Rev. D {\bf 78}, 116018 (2008).
\bibitem{A411} G. Altarelli, F. Feruglio and C. Hagedorn, JHEP \textbf{0803}, 052 (2008).
\bibitem{A412} M. Hirsch, S. Morisi and J. W. F. Valle, Phys. Rev. D {\bf 78}, 093007 (2008).
\bibitem{A413}E. Ma, Phys. Lett. B {\bf 671}, 366 (2009).
\bibitem{A414}G. Altarelli and D. Meloni, J. Phys. G {\bf 36}, 085005 (2009).
\bibitem{A415} Y. Lin, Nucl. Phys. B {\bf 813}, 91 (2009).
\bibitem{A416} Y. H. Ahn and C. S. Chen, Phys. Rev. D {\bf 81}, 105013 (2010).
\bibitem{A417} J. Barry and W. Rodejohanny, Phys. Rev. D {\bf 81}, 093002 (2010).
\bibitem{A418} G. J. Ding and D. Meloni, Nucl. Phys. B {\bf 855}, 21 (2012).
\bibitem{A419}\textbf{ L. Lavoura, H. Kuhbock, Eur. Phys.J. C55 (2008) 303, arXiv:0711.0670 [hep-ph].}
\bibitem{dlsh}  P. V. Dong, L. T. Hue, H. N. Long, D. V. Soa, Phys. Rev. D 81,
053004 (2010).


\bibitem{A51}  A. Datta, F. S.  Ling, P. Ramond, Nucl. Phys. B671 (2003) 383-400.
\bibitem{A52}C. Luhn, S. Nasri, P. Ramond, J. Math. Phys. 48, 123519 (2007), arXiv:0709.1447 [hep-th].
\bibitem{A53} Y. Kajiyama, M. Raidal, A. Strumia, Phys. Rev. D76, 117301(2007),  arXiv:0705.4559 [hep-ph].
\bibitem{A54} C. Luhn, P. Ramond, J. Math. Phys.49, 053525 (2008), arXiv:0803.0526 [hep-th].
\bibitem{A55}L. L. Everett, A. J. Stuart, Phys. Rev. D79, 085005 (2009).
\bibitem{A56} A. Adulpravitchai, A. Blum, W. Rodejohann, \emph{Golden Ratio Prediction for Solar Neutrino Mixing},
  arXiv: 0903.0531 [hep-ph].
\bibitem{A57} C. S.  Chen, T.  W. Kephart, T. C. Yuan, \emph{An $A_5$ Model of Four Lepton Generations}, arXiv:1011.3199 [hep-ph].
\bibitem{A58} I. K. Cooper, S. F. King,  A. J. Stuart, \emph{A  Golden $A_5$ Model of Leptons with a Minimal NLO Correction}, arXiv:1212.1066 [hep-ph].
\bibitem{A59} C. S.  Chen, T. W. Kephart, T. C.  Yuan, \emph{ Binary Icosahedral Flavor Symmetry for Four Generations of Quarks and Leptons},
  arXiv:1110.6233 [hep-ph].
\bibitem{A510} K. Hashimoto, H. Okada, \emph{Lepton Flavor Model and Decaying Dark Matter in The Binary Icosahedral Group Symmetry}, arXiv:1110.3640 [hep-ph].
\bibitem{A511} G. J.  Ding, L. L. Everett, A. J. Stuart, \emph{Golden Ratio Neutrino Mixing and $A_5$ Flavor Symmetry}, arXiv:1110.1688 [hep-ph].
\bibitem{A512} F. Feruglio, A. Paris, JHEP 1103,  101 (2011).
\bibitem{A513} L. L. Everett, A. J. Stuart, \emph{The Double Cover of the Icosahedral Symmetry Group and Quark Mass Textures}, arXiv:1011.4928 [hep-ph].



\bibitem{S31} L. Wolfenstein, Phys. Rev. D {\bf 18}, 958 (1978).
\bibitem{S32}S. Pakvasa and H. Sugawara, Phys. Lett. B {\bf 73}, 61 (1978).
\bibitem{S33}S. Pakvasa and H. Sugawara, Phys. Lett. B {\bf 82}, 105 (1979).
 \bibitem{S34} E. Durman and H. S. Tsao, Phys. Rev. D {\bf 20}, 1207 (1979).
 \bibitem{S35} Y. Yamanaka, H. Sugawara, and S. Pakvasa, Phys. Rev. D {\bf 25}, 1895 (1982)
 \bibitem{S36} K. Kang, J. E. Kim, and P. Ko, Z. Phys. C {\bf 72}, 671 (1996).
 \bibitem{S37} H. Fritzsch and Z. Z. Xing, Phys. Lett. B {\bf 372}, 265 (1996).
 \bibitem{S38} K. Kang, S. K. Kang, J. E. Kim, and P. Ko, Mod. Phys. Lett. A {\bf 12}, 1175 (1997).
 \bibitem{S39} M. Fukugita, M. Tanimoto, and T. Yanagida, Phys. Rev. D {\bf 57}, 4429 (1998).
 \bibitem{S310} H. Fritzsch and Z. Z. Xing, Phys. Lett. B {\bf 440}, 313 (1998)
 \bibitem{S311} Y. Koide, Phys. Rev. D {\bf 60}, 077301 (1999).
 \bibitem{S312} H. Fritzsch and Z. Z. Xing, Phys. Rev. D {\bf 61}, 073016 (2000).
 \bibitem{S313} M. Tanimoto, Phys.Lett. B {\bf 483}, 417 (2000).
 \bibitem{S314} G. C. Branco and J. I.Silva-Marcos, Phys. Lett. B {\bf 526}, 104 (2002).
 \bibitem{S315} M. Fujii, K. Hamaguchi and T. Yanagida, Phys. Rev. D {\bf 65}, 115012 (2002).
 \bibitem{S316} J. Kubo, A. Mondragon, M. Mondragon and E. Rodriguez-Jauregui, Prog. Theor. Phys. {\bf 109}, 795 (2003).
 \bibitem{S317} J. Kubo, A. Mondragon, M. Mondragon and E. Rodriguez-Jauregui, Prog. Theor. Phys. {\bf 114}, 287(E) (2005).
 \bibitem{S318} P. F. Harrison and W. G. Scott, Phys. Lett. B {\bf 557}, 76 (2003).
 \bibitem{S319} S.-L. Chen, M. Frigerio, and E. Ma, Phys. Rev. D {\bf 70}, 073008 (2004).
 \bibitem{S320} H. Fritzsch and Z. Z. Xing, Phys. Lett. B {\bf 598}, 237 (2004).
 \bibitem{S321} F. Caravaglios and S. Morisi, arXiv: hep-ph/0503234.
 \bibitem{S322} W. Grimus and L. Lavoura, JHEP {\bf 0508}, 013 (2005).
 \bibitem{S323} R. N. Mohapatra, S. Nasri and H. B. Yu, Phys. Lett. B {\bf 639}, 318 (2006).
 \bibitem{S324} R. Jora, S. Nasri and J. Schechter, Int. J. Mod. Phys. A {\bf 21}, 5875 (2006).
 \bibitem{S325} J. E. Kim and J.-C. Park, JHEP {\bf 0605}, 017 (2006).
 \bibitem{S326} Y. Koide, Eur. Phys. J. C {\bf 50}, 809 (2007).
 \bibitem{S327} A. Mondragon, M. Mondragon, and E. Peinado, Phys. Rev. D {\bf 76}, 076003 (2007).
 \bibitem{S328} A. Mondragon, M. Mondragon, and E. Peinado, AIP Conf. Proc. {\bf 1026}, 164 (2008).
 \bibitem{S329} M. Picariello, Int. J. Mod. Phys. A {\bf 23}, 4435 (2008).
 \bibitem{S330} C. Y. Chen and L. Wolfenstein, Phys. Rev. D {\bf 77}, 093009 (2008).
 \bibitem{S331} R. Jora, J. Schechter and M. Naeem Shahid, Phys. Rev. D {\bf 80}, 093007 (2009).
 \bibitem{S332} R. Jora, J. Schechter and M. Naeem Shahid, Phys. Rev.  {\bf 82}, 079902(E) (2010).
\bibitem{S333}D. A. Dicus, S. F. Ge and W. W. Repko, Phys. Rev. D {\bf 82}, 033005 (2010).
\bibitem{S334} Z. Z. Xing, D. Yang and S. Zhou, Phys. Lett. B {\bf 690}, 304 (2010).
\bibitem{S335} R. Jora, J. Schechter and M. N. Shahid, Phys. Rev. D {\bf 82}, 053006 (2010).
\bibitem{S336} S. Dev, S. Gupta and R. R. Gautam, Phys. Lett. B {\bf 702}, 28 (2011).
\bibitem{S337} D. Meloni, S. Morisi and E.  Peinado, J. Phys. G {\bf 38}, 015003 (2011).
\bibitem{S338} G. Bhattacharyya, P. Leser and H. Pas, Phys. Rev. D {\bf 83}, 011701(R) (2011).
\bibitem{S339} T. Kaneko and H. Sugawara, Phys. Lett. B {\bf 697}, 329 (2011).
\bibitem{S340} S. Zhou, Phys. Lett. B {\bf 704}, 291 (2011).
\bibitem{S341} F. Gonz$\mathrm{\acute{a}}$lez Canales \emph{et.al}, Quark sector of S3 models: classification and comparison with
experimental data, arXiv:1304.6644 [hep-ph].
\bibitem{S342} E. Ma, B. $\mathrm{Meli\acute{c}}$,  \emph{Updated $S_3$ model of quarks},
 Phys. Lett.\textbf{ B 725} (2013) 402.

\bibitem{S41} R. N. Mohapatra, M. K. Parida, G. Rajasekaran, Phys. Rev. D \textbf{69}, 053007 (2004).
\bibitem{S42} C. Hagedorn, M. Lindner, and R. N. Mohapatra, JHEP \textbf{0606}, 042 (2006).
\bibitem{S43} E. Ma, Phys. Lett. B \textbf{632}, 352 (2006).
\bibitem{S44} H. Zhang, Phys. Lett. B \textbf{655}, 132 (2007).
\bibitem{S45} Y. Koide, JHEP \textbf{0708}, 086 (2007).
\bibitem{S46} H. Ishimori, Y. Shimizu, and M. Tanimoto, Prog. Theor. Phys. \textbf{121}, 769 (2009).
\bibitem{S47}F. Bazzocchi, L. Merlo, and S. Morisi, Nucl. Phys. B {\bf 816},204 (2009).
\bibitem{S48} F. Bazzocchi and S. Morisi, Phys. Rev. D \textbf{80} 096005 (2009).
\bibitem{S49} G. Altarelli and F. Fergulio, Rev. Mod. Phys. {\bf 82}, 2701 (2010).
\bibitem{S410} G. J. Ding, Nucl. Phys. B \textbf{827}, 82 (2010).
\bibitem{S411}Y. H. Ahn, S. K. Kang, C. S. Kim, and T. P. Nguyen, Phys.Rev. D {\bf 82}, 093005 (2010).
\bibitem{S412}Y. Daikoku and H. Okada, arXiv:1008.0914 [hep-ph].
\bibitem{S413}H. Ishimori, Y. Shimizu, M. Tanimoto, and A. Watanabe, Phys. Rev. D {\bf 83} 033004 (2011).
\bibitem{S414} H. Ishimori and M. Tanimoto, Prog. Theor. Phys. {\bf 125}, 653 (2011).
\bibitem{S415}R. Z. Yang and H. Zhang, Phys. Lett. B {\bf 700} 316, (2011).
\bibitem{S416}S. Morisi and E. Peinado, Phys. Lett. B {\bf 701}, 451 (2011).
\bibitem{S417}S. Morisi, K.M. Patel, and E. Peinado, Phys. Rev. D {\bf 84}, 053002 (2011).
\bibitem{S418}L. Dorame, S. Morisi, E. Peinado, J. W. F. Valle and Alma D. Rojas, Phys. Rev. D \textbf{86}, 056001 (2012), [arXiv:1203.0155
(hep-ph)].
\bibitem{S419}D. Hernandez, A. Yu. Smirnov, Phys. Rev D \textbf{86}, 053014 (2012), [arXiv:1204.0445 (hep-ph)].
\bibitem{S420}Z. H.  Zhao, Phys. Rev. D \textbf{86}, 096010 (2012), [arXiv:1207.2545 (hep-ph)].
\bibitem{S421}R. Krishnan, P. F. Harrison, W. G. Scott, \emph{Simplest Neutrino Mixing from S4 Symmetry}, arXiv:1211.2000 [hep-ph].
\bibitem{S422}R. Krishnan, \emph{A Model for Large $\theta_{13}$ Constructed using the Eigenvectors of the$S_4$ Rotation Matrices}, arXiv: 1211.3364 [hep-ph].
\bibitem{S423}I.  de M.  Varzielas, L. Lavoura, \emph{Flavour models for TM1 lepton mixing} , arXiv: 1212.3247 [hep-ph]
\bibitem{S424} W. Grimus, \emph{Discrete symmetries, roots of unity, and lepton mixing}, arXiv:1301.0495 [hep-ph].
\bibitem{S425}S. F. King, C. Luhn, \emph{Neutrino Mass and Mixing with Discrete Symmetry}, arXiv:1301.1340 [hep-ph].
\bibitem{S426}R. G. Felipe, H. Serodio, J.  P. Silva, \emph{Models with three Higgs doublets in the triplet representations of $A_4$ or $S_4$}, arXiv:1302.0861[hep-ph].
\bibitem{S427} Y. Daikoku, H. Okada, \emph{Phenomenology of $S_4$ Flavor Symmetric extra $\mathrm{U(1)}$ model}, arXiv:1303.7056 [hep-ph].
\bibitem{S428}F. Feruglio, C. Hagedorn, R. Ziegler,\emph{ $S_4$ and CP in a SUSY
Model}, arXiv:1303.7178 [hep-ph].
\bibitem{S429}Ch. Luhn, \emph{Trimaximal TM1 neutrino mixing in $S_4$ with spontaneous CP violation}, arXiv:1306.2358 [hep-ph].



\bibitem{D41} P.  H. Frampton, T. W. Kephart, Int. J. Mod. Phys. A10, 4689 (1995), arXiv:hep-ph/9409330.
\bibitem{D42}P.  H. Frampton, T. W. Kephart, Phys. Rev. D64 (2001) 086007, arXiv:hep-th/0011186.
\bibitem{D43}W. Grimus, L. Lavoura, Phys.Lett. B572, 189 (2003), arXiv:hep-ph/0305046.
\bibitem{D44} W. Grimus, L. Lavoura, \emph{Models of maximal atmospheric neutrino mixing and leptogenesis}, arXiv:hep-ph/0405261.
\bibitem{D45}W. Grimus, A.S. Joshipura, S. Kaneko, L. Lavoura, M. Tanimoto, JHEP 0407 (2004) 078, arXiv:hep-ph/0407112.
 \bibitem{D46}M. Frigerio, S. Kaneko, E.Ma, M. Tanimoto, Phys. Rev. D 71 (2005) 011901, arXiv:hep-ph/0409187.
 \bibitem{D47} K. S. Babu, J. Kubo, Phys.Rev. D71 (2005) 056006, arXiv:hep-ph/0411226.
 \bibitem{D48} M. Honda, R. Takahashi, M. Tanimoto, JHEP 0601 (2006) 042, arXiv:hep-ph/0510018.
 \bibitem{D49} H. Ishimori \emph{et al.}, Phys. Lett. B 662, 178 (2008), arXiv:0802.2310 [hep-ph].
 \bibitem{D410} H. Abe, K-S. Choi, T. Kobayashi b, H. Ohki, Nucl.  Phys.  B 820, 317 (2009).
 \bibitem{D411} T. Araki \emph{et al.}, Nucl. Phys.  B 805, 124 (2008).
 \bibitem{D412} A. Adulpravitchai, A. Blum, C. Hagedorn,  JHEP 0903 (2009) 046, arXiv: 0812.3799 [hep-ph].



\bibitem{D51} C. Hagedorn, M. Lindner, F. Plentinger, Phys. Rev. D74 (2006) 025007, arXiv:hep-ph/0604265.
\bibitem{D52} T. Kobayashi, H. P. Nilles, F. Pl$\mathrm{\ddot{o}}$ger, S. Raby and M. Ratz, Nucl. Phys.\textbf{ B 768}:135,2007, arXiv:hep-ph/0611020.



\bibitem{Tp1} F. Feruglio, C. Hagedorn, Y. Lin and L. Merlo, Nucl. Phys. B \textbf{775}, 120 (2007).
\bibitem{Tp2} M. C. Chen and K. T. Mahanthappa, Phys. Lett. B \textbf{652}, 34 (2007).
\bibitem{Tp3} P. H. Frampton and T. W. Kephart, JHEP \textbf{0709}, 110 (2007).
\bibitem{Tp4} G. J. Ding, Phys. Rev. D \textbf{78} 036011 (2008).
\bibitem{Tp5} G. J. Ding, Phys. Rev. D 78, 036011 (2008), arXiv:0803.2278 [hep-ph].
\bibitem{Tp6} P. H. Frampton, S. Matsuzaki, Phys. Lett. B \textbf{679}, 347 (2009).
\bibitem{Tp7} P. H. Frampton and S. Matsuzaki, Phys. Lett. B \textbf{679} 347 (2009)
\bibitem{Tp8} D. A. Eby, P. H. Frampton and S. Matsuzaki, Phys. Lett. B \textbf{671}, 386 (2009), arXiv:0810.4899[hep-ph].
\bibitem{Tp9} C. M. Ho and T. W. Kephart, Phys. Lett. B 687, 201 (2010) [arXiv:1001.3696 [hep-ph]].
\bibitem{Tp10} P. H. Frampton, C. M. Ho, T. W. Kephart and S. Matsuzaki, Phys. Rev. D 82, 113007 (2010), arXiv:1009.0307 [hep-ph].
\bibitem{Tp11} D. A. Eby, P. H. Frampton, X. -G. He and T. W. Kephart, Phys. Rev. D 84, 037302 (2011), arXiv:1103.5737 [hep-ph].
\bibitem{Tp12} P. H. Frampton, C. M. Ho, T. W. Kephart, Phys. Rev. D \textbf{89} (2014) 027701, arXiv:1305.4402 [hep-ph].


\bibitem{T71} C. Luhn, S. Nasri and P. Ramond, Phys. Lett. B 652, 27 (2007), arXiv:0706.2341 [hep-ph].
\bibitem{T72} C. Hagedorn, M. A. Schmidt and A. Y. .Smirnov, Phys. Rev. D 79, 036002 (2009), arXiv:0811.2955 [hep-ph].
\bibitem{T73} Q. -H. Cao, S. Khalil, E. Ma and H. Okada, Phys. Rev. Lett. 106, 131801 (2011), arXiv:1009.5415 [hep-ph].
\bibitem{T74} Q. -H. Cao, S. Khalil, E. Ma and H. Okada, Phys. Rev. D 84, 071302 (2011), arXiv:1108.0570 [hep-ph].
\bibitem{T75} H. Ishimori, S. Khalil and E. Ma, Phys. Rev. D 86, 013008 (2012), arXiv:1204.2705 [hep-ph].
\bibitem{T76} Andrzej J. Buras, Fulvia De Fazio, Jennifer Girrbach, \emph{331 models facing new $b\to s\mu^+\mu^-$ data}, arXiv:1311.6729 [hep-ph].
\bibitem{T77} Fulvia De Fazio, \emph{Quark flavour observables in 331 models in the flavour precision era}, arXiv:1310.4614 [hep-ph].




\bibitem{331m1} J. W. F. Valle, M. Singer, Phys. Rev. D \textbf{28} (1983) 540.
\bibitem{331m2} F. Pisano and V. Pleitez, Phys. Rev.  D {\bf 46}, 410 (1992).
\bibitem{331m3} P. H. Frampton, Phys. Rev. Lett. {\bf 69}, 2889 (1992).
\bibitem{331m4} R. Foot, O. F. Hernandez, F. Pisano and V. Pleitez, Phys. Rev. D {\bf 47}, 4158 (1993).
\bibitem{331m5}J. C. Montero, F. Pisano and V. Pleitez, Phys. Rev. \textbf{D} 47, 2918 (1993).



\bibitem{331r1} M. Singer, J. W. F. Valle and J. Schechter, Phys. Rev. D {\bf 22}, 738 (1980).
\bibitem{331r2} R. Foot, H. N. Long and Tuan A. Tran, Phys. Rev. D {\bf 50}, R34 (1994), arXiv: 9402243 [hep-ph].
\bibitem{331r3} J. C. Montero, F. Pisano and V. Pleitez, Phys. Rev. D {\bf 47}, 2918 (1993).
\bibitem{331r4} H. N. Long, Phys. Rev. D {\bf 54}, 4691 (1996).
\bibitem{331r5} H. N. Long, Phys. Rev. D {\bf 53}, 437 (1996).
\bibitem{331r6} H. N. Long,  Mod. Phys. Lett. A {\bf 13}, 1865 (1998), [arXiv: hep-ph/9711204].


\bibitem{e3311} W. A. Ponce, Y. Giraldo and L. A. Sanchez, Phys. Rev. D {\bf 67}, 075001 (2003).
\bibitem{e3312} P. V. Dong, H. N. Long, D. T. Nhung and D. V. Soa, Phys. Rev. D {\bf 73}, 035004 (2006).
\bibitem{e331v1} P. V. Dong, D. T. Huong, Tr. T. Huong, H. N. Long, Phys.Rev.\textbf{D74}, 053003 (2006).
\bibitem{e331v2} P. V. Dong, H. N. Long, D. V. Soa, Phys.Rev.\textbf{D75}, 073006 (2007).
\bibitem{e3313} P. V. Dong, H. T. Hung and H. N. Long, Phys. Rev. \textbf{D 86}, 033002 (2012),
 arXiv:1205.5648 [hep-ph].
\bibitem{e3314} P. V. Dong and H. N. Long, Adv. High Energy Phys. {\bf 2008}, 739492 (2008),
arXiv:0804.3239 [hep-ph].



\bibitem{dlshA4} P. V. Dong, L. T. Hue, H. N. Long
and D. V. Soa, Phys. Rev. D {\bf 81}, 053004 (2010).


\bibitem{dlsvS4} P. V. Dong, H. N. Long, D. V. Soa, and V. V. Vien,   Eur.
Phys. J. C \textbf{71}, 1544 (2011), arXiv:1009.2328 [hep-ph].


\bibitem{dlnvS3} P. V. Dong, H. N. Long, C. H. Nam, and V. V. Vien,  Phys. Rev.
 D85, 053001(2012), arXiv: 1111.6360 [hep-ph].

 \bibitem{vlD4} V. V. Vien and H. N. Long, Int. J.  Mod. Phys. A \textbf{28} (2013) 1350159,
 arXiv: 1312.5034 [hep-ph].

\bibitem{CPvio} Y. H. Ahn, S. K. Kang, C. S. Kim, \emph{Spontaneous CP Violation in $A_4$
Flavor Symmetry and Leptogenesis},  arXiv:1304.0921 [hep-ph].



 \bibitem{clong} D. Chang and H. N. Long, Phys. Rev. {\bf D 73}, 053006 (2006).
 \bibitem{TBM2}\textbf{ C.  H. Albright, W.  Rodejohann, Eur. Phys. J. C 62 (2009) 599, arXiv:0812.0436 [hep-ph].}
\bibitem{MaximalCP} D. Marzocca, S. T. Petcov, A. Romanino, M. C. Sevilla, J. High E. Phys. 05 (2013) 073, arXiv:1302.0423 [hep-ph].

 \bibitem{Jarlskog} X. Zhang, B-Q. Ma, Phys. Lett. B 713 (2012) 202-205, arXiv: 1203.2906 [hep-ph].


\bibitem{Wein} C. Weinheimer, Proceedings of the 20th International Conference on
Neutrino Physics and Astrophysics, Neutrino 2002 (Munich, Germany) May 25-30, 2002,
 Nucl. Phys. Proc. Suppl. 118, 388 (2003), arXiv: 0209556 [astro-ph].
\bibitem{Lobashev} V. Lobashev et al, Nucl. Phys. Proc. Suppl. 91 (2001) 280.



 \bibitem{Tegmark} M. Tegmark et al, Phys. Rev. D 69 (2004) 103501, arXiv: 0310723 [astro-ph].


 \bibitem{Weiler} T. J.  Weiler (2013), \emph{Oscillation and Mixing Among the Three Neutrino Flavors},
arXiv:1308.1715 [hep-ph].

\textbf{\bibitem{betdecay1} W. Rodejohann, Int. J. Mod. Phys. E
20, 1833 (2011), arXiv: 1106.1334 [hep-ph].}
\textbf{\bibitem{betdecay2} M. Mitra, G. Senjanovic, F. Vissani,
Nucl. Phys. B 856, 26 (2012), arXiv:1108.0004 [hep-ph]}
\textbf{\bibitem{betdecay3}S. M. Bilenky, C. Giunti,
\emph{Neutrinoless double-beta decay. A brief review},
arXiv:1203.5250 [hep-ph]. } \textbf{\bibitem{betdecay4} W.
Rodejohann, \emph{Neutrinoless double beta decay and neutrino
physics}, arXiv:1206.2560 [hep-ph]. } \textbf{\bibitem{betdecay5}
A.  Merle, Int. J. Mod. Phys. D 22, 1330020 (2013),
arXiv:1302.2625 [hep-ph].} \textbf{\bibitem{betdecay6} J. D.
Vergados, H. Ejiri and F. Simkovic, \emph{Theory of neutrinoless
double beta decay}, arXiv:1205.0649 [hep-ph]. }
\textbf{\bibitem{expbet1} C. Kraus \emph{et al.}, Eur. Phys. J. C
40 (2005) 447, arXiv: 0412056 [hep-ex].} \textbf{\bibitem{expbet2}
V. Aseev \emph{et al.}, Phys. Rev. D 84 (2011) 112003,
arXiv:1108.5034 [hep-ex]. }
\bibitem{DongLongdh} P. V. Dong and H. N. Long, Eur. Phys. J. C 42, 325 (2005).
\bibitem{DongHLT} P. V. Dong, V. T. N. Huyen , H. N. Long, and H. V. Thuy,
\emph{Gauge Boson Mixing in the 3-3-1 Models with Discrete
Symmetries}, Adv. High Energy Phys. {\bf 2012}, 715038 (2012).

\bibitem{S3quark} A. E. C.  Hernandez, R. Martinez, J.  Nisperuza,
\emph{$S_3$ flavour symmetry breaking scheme for understanding the quark mass and mixing pattern
 in $SU(3)_C\otimes SU(3)_L\otimes U(1)_X$ models}, arXiv:1401.0937 [hep-ph].


{\bf \bibitem{stu}  P. H. Frampton and M. Harada, Phys. Rev. D 58
(1998) 095013;
 H. N. Long and T. Inami,  Phys. Rev. D 61 (2000) 075002.}


 \end{thebibliography}
\end{document}